\documentclass[prd,twocolumn,preprintnumbers,nofootinbib,showpacs]{revtex4-1}
\usepackage{}
\usepackage{bm}
\usepackage{amsmath,amssymb}
\usepackage{rotating}
\usepackage{graphicx}
\usepackage{subfigure}

\newcommand{\GeV}{\;\text{GeV}}
\newcommand{\beqn}{\begin{eqnarray}}
\newcommand{\eeqn}{\end{eqnarray}}

\usepackage{color}
\usepackage{float}

\begin{document}

\title{Inverse Magnetic Catalysis in the three-flavor NJL model with axial-vector interaction}

\author{Lang Yu$^{1}$}
%\email{yulang@mail.ihep.ac.cn}
\author{Jos Van Doorsselaere$^{2}$}
%\email{haoliu@mail.ihep.ac.cn}
\author{Mei Huang$^{1,3}$}
%%\email{huangm@mail.ihep.ac.cn}
\affiliation{$^1$ Institute of High Energy Physics, Chinese Academy of Sciences,
Beijing 100049, China}
\affiliation{$^2$ Laboratoire de Mathemathique et Physique Theorique, Universite de Tours, 37000 Tours, France}
\affiliation{$^3$ Theoretical Physics Center for Science Facilities,
Chinese Academy of Sciences,
Beijing 100049, China}
\date{\today}

\begin{abstract}
In this paper we explore the chiral phase transition in QCD within the three-flavor Nambu-Jona-Lasinio (NJL) model with a negative coupling constant in the isoscalar axial-vector channel, which is associated with a polarized instanton--anti-instanton molecule background. The QCD phase diagram described in this scenario shows a new first order phase transition around the transition temperature $T_c$ toward a phase without chiral condensates, but with nontrivial dynamic chiral chemical potentials for the light quarks, spontaneously giving rise to local $\mathcal {CP}$ violation and local chirality imbalance. The corresponding critical temperature $T_{5c}$ for this phase transition decreases with the magnetic field and it gives a natural explanation to the inverse magnetic catalysis effect for light quarks when incorporating a reasonable value of the coupling constant in the isoscalar axial-vector channel. Furthermore,  when the isoscalar axial-vector interaction is dominant in light quark sector and suppressed in strange quark sector,  it is found that  there is no inverse magnetic catalysis for strange quark condensate, which agrees with lattice results.
\end{abstract}
\maketitle

\section{Introduction}

Investigation of the QCD phase structure  in the presence of strong
external magnetic fields has become a major topic in both theoretical and experimental research into the physics of strongly interacting matter.
This topic is of paramount importance to understand the
phenomenology of noncentral heavy ion collisions at the Relativistic Heavy Ion Collider (RHIC) and
the Large Hadron Collider (LHC), in which a strong magnetic field reaching up to $\sqrt{eB}\sim(0.1-1.0)$ GeV ~\cite{Skokov:2009qp,Voronyuk:2011jd,Bzdak:2011yy,Deng:2012pc} can be
generated. In addition, strong magnetic fields could also have existed in the strong and electroweak phase
transition~\cite{Vachaspati:1991nm,Enqvist:1993np} of the early Universe, and exist in compact stars like magnetars~\cite{Duncan:1992hi}.

The breaking and restoration of chiral symmetry, which is described by the behavior of the quark condensate,
is one of the most intriguing nonperturbative aspects of QCD. Therefore, it is of great
interest to speculate the effect of magnetic fields on the behavior of the chiral
condensate in QCD at zero and finite temperatures. Since the 1990's, a related phenomenon known as magnetic catalysis has been
recognized ~\cite{Klevansky:1989vi, Klimenko:1990rh,Gusynin:1995nb, Shovkovy:2012zn}. It refers to an enhancement
of the quark condensate and thus an increase of the chiral transition temperature $T_c$ under the magnetic field.
This is agreed by most of earlier low-energy effective models and approximations to QCD~
\cite{Klevansky:1989vi, Klimenko:1990rh, Gusynin:1995nb, Shushpanov:1997sf, Agasian:1999sx, Alexandre:2000yf,
Agasian:2001hv, Cohen:2007bt, Gatto:2010qs, Gatto:2010pt, Mizher:2010zb, Kashiwa:2011js,
Avancini:2012ee, Andersen:2012dz, Scherer:2012nn} as well as lattice QCD simulations~\cite{Buividovich:2008wf,
Braguta:2010ej, D'Elia:2010nq, D'Elia:2011zu,Ilgenfritz:2012fw}
in the past twenty years. However, recently, a lattice group~\cite{Bali:2011qj,Bali:2012zg} revealed
surprising results that the transition temperature $T_c$ decreases as a function of the external magnetic field,
and the chiral condensate shows a nonmonotonic behavior as a function of the external magnetic field in the crossover
region. This prediction is in contrast to the majority of previous calculations,
and the partly decreasing behavior of the chiral condensate with the increasing $B$ near $T_c$, causing a decreasing dependence of $T_c$ on $B$, is called inverse magnetic catalysis.

There are several recent studies~\cite{Fukushima:2012kc, Kojo:2012js, Bruckmann:2013oba, Chao:2013qpa, Fraga:2013ova, Ferreira:2014kpa, Farias:2014eca, Yu:2014sla, Andersen:2014oaa, Ferrer:2014qka, Feng:2014bpa} discussing the origin of the phenomenon of the decreasing behavior of the chiral critical temperature with increasing $B$ and the inverse magnetic catalysis around $T_c$. For example, the magnetic inhibition~\cite{Fukushima:2012kc},
the mass gap in the large $N_c$ limit~\cite{Kojo:2012js}, the contribution of sea quarks~\cite{Bruckmann:2013oba}, a running  scalar coupling parameter dependent on the magnetic field intensity~\cite{Ferreira:2014kpa,Farias:2014eca} and an antiscreening effect of the color charges for quarks~\cite{Ferrer:2014qka, Ferrer:2013noa}, etc are proposed to understand this puzzle. Particularly, a very natural and competitive mechanism
attributes the inverse magnetic catalysis to the local chirality imbalance induced by the nontrivial topological
gluon configuration, arising from a sphaleron transition \cite{Chao:2013qpa}
or the instanton--anti-instanton molecule pairing \cite{Yu:2014sla}.

As discussed in Ref.~\cite{Yu:2014sla},
the chirality imbalance, which is associated with the violation of the $\mathcal {P}$ and $\mathcal {CP}$ symmetry,
is induced by the nonzero topological charge $Q_T$ through the axial anomaly of QCD
\beqn
\Delta N_5=\int{d^4 x}\partial_{\mu}j^{\mu}_5=-2N_fQ_T,
\label{eq:DN5}
\eeqn
where $N_f$ is the number of flavors, $\Delta N_5=N_5(t=+\infty)-N_5(t=-\infty)$,
with $N_5=N_R-N_L$ denoting the number difference between right- and left-hand quarks, and
$j^{\mu}_5=\bar\psi \gamma^{\mu} \gamma^5 \psi$ denotes the isospin singlet axial vector current.
A consequence of this violation is the existence of two kinds of local domains with the same quantum numbers but opposite
net topological charges, which will lead to the generation of local chirality imbalances but zero
average chirality, as well as the local $\mathcal {P}$ and $\mathcal {CP}$ violation. The modification of the QCD phase diagram by the chirality imbalance has been studied in some Refs.~\cite{Fukushima:2010fe, Chernodub:2011fr, Gatto:2011wc}. Especially,
the recent observation of charge azimuthal correlations at RHIC and LHC~\cite{Abelev:2009ac,Abelev:2009ad,Abelev:2012pa}
may be resulting from the chiral magnetic effect (CME) with local $\mathcal {P}$ and $\mathcal {CP}$ violation, which
is an interesting combined effect of both the strong magnetic field and the nontrivial topological
gluon configuration of the quark-gluon plasma. Based on above discussions, the enhancement of the chirality imbalance
by magnetic fields, which coincides with the lattice results in Ref.~\cite{Buividovich:2009wi}, will naturally result in a decreasing chiral critical temperature, since it
destroys the pairing between the left-handed quark (antiquark) and the right-handed antiquark (quark).

In Ref.~\cite{Yu:2014sla}, a mechanism has been presented to generate local
chirality imbalance based on the instanton--anti-instanton ($I\bar{I}$) molecule picture
~\cite{Ilgenfritz:1988dh, Ilgenfritz:1994nt, Schafer:1994nv, Schafer:1996wv, Zhang:2012rv}, which is regarded as
one effective mechanism responsible for nonperturbative properties of QCD in the region
$T\simeq T_c - 2T_c$~\cite{Schafer:1994nv,Schafer:1996wv, Zhang:2012rv}. By using an unconventional repulsive
iso-scalar axial-vector interaction in  a two flavor Nambu-Jona-Lasinio (NJL) model ~\cite{Schafer:1994nv}, we find that
a dynamical chiral chemical potential related to the local chirality imbalance
is induced spontaneously at the temperatures near $T_c$. It is also found that
the increasing magnetic field helps to lower the critical temperature due to the
appearance of the local chirality. Moreover, since the local chirality imbalance can only be produced
around $T_c$, it gives a reasonable explanation for why inverse magnetic catalysis only appears
at the temperatures around $T_c$, while magnetic catalysis still occurs at zero and low temperatures.

However, the lattice result in Ref.\cite{Bali:2012zg} shows that the strange quark condensate does not
exhibit inverse magnetic catalysis, but simply increases with magnetic field at all temperatures.
In this paper, we extend our analysis and calculations to the 2+1 flavors, and investigate
the corresponding effects of the axial-vector interaction on both the local chirality imbalance and
the chiral critical temperatures for u, d and s quarks. The paper is organized as follows.
In Sec. II, we give a general description of the 2+1 flavor NJL model and formalism with considering
the repulsive axial-vector interactions stemming from the interacting $I\bar{I}$ molecule model (IIMM).
In Sec. III, we will discuss the main results of the numerical calculations. Finally, our conclusions and
perspectives are presented In Sec. IV.

\section{Model and Formalism}

In this section, we present the three flavor NJL model~\cite{Nambu:1961tp,
Nambu:1961fr, Bernard:1987gw, Bernard:1987sg, Klimt:1989pm, Vogl:1991qt, Lutz:1992dv, Klevansky:1992qe, Hatsuda:1994pi, Buballa:2003qv} with adding the vector and
axial-vector interaction terms. The Lagrangian density of our model in the presence of an
external magnetic field is given by
\beqn
 \mathcal{L} & = & \bar\psi (i\gamma_\mu  D^\mu-\hat{m})\psi+\mathcal{L}_{sym}
 +\mathcal{L}_{det}-\mathcal{L}_{VA}\;,
\label{eq:L:basic}
\eeqn
where $\psi=(u,d,s)^T$ corresponds to the quark field of three flavors, and
$\hat{m}= \text{diag} (m_u,\, m_d,\, m_s)$ is the corresponding current mass matrix.
The covariant derivative, $ D_{\mu}=\partial_{\mu}-i q_f A_{\mu}$, couples quarks to an external
magnetic field $\bm{B}=(0,0,B)$ along the positive $z$ direction, via a background Abelian gauge field
$A^{\mu}=(0,0,Bx,0)$. And $q_f$ is defined as the electric charge of the quark field with flavor $f$.
$\mathcal{L}_{sym}$ and $\mathcal{L}_{det}$ are given by
\beqn
 \mathcal{L}_{sym} & = & \frac{G_S}{2}\sum_{a=0}^8\left[\left(\bar\psi\lambda^a\psi\right)^2
 +\left(\bar\psi\lambda^a i \gamma^5\psi\right)^2\right], \\
  \mathcal{L}_{det} & = & -K\left\{\det\left[\bar\psi(1+\gamma^5)\psi\right]
  +\det\left[\bar\psi(1-\gamma^5)\psi\right]\right\}\;,\nonumber\\
\label{eq:L:symdet}
\eeqn
where $\lambda^a$ are the Gell-Mann matrices in flavor space ($\lambda^0=\sqrt{2/3}I$)
and the determinant is in flavor space also. The $\mathcal{L}_{sym}$ term corresponds to the usual
four-fermion interactions of scalar and pseudoscalar channels which
respect $SU(3)_V\otimes SU(3)_A\otimes U(1)_V \otimes U(1)_A$ symmetry. The $\mathcal{L}_{det}$ term
corresponds to the 't Hooft six-fermion determinant interactions~\cite{'tHooft:1976fv} which break $U(1)_A$ symmetry.
As for the $\mathcal{L}_{VA}$ term, it represents the four-fermion interactions of vector and axial-vector
channels under the invariance of $SU(3)_V\otimes SU(3)_A\otimes U(1)_V \otimes U(1)_A$ symmetry.
The last two terms are added to introduce chiral interactions, which are chosen such that they correspond to
effective chirally asymmetric interactions with an instanton background.

The connection between instantons and chiral symmetry breaking is well known. Essentially it is a consequence of an
index theorem that shows how a topologically non-trivial gauge configuration --the instanton-- gives rise to an asymmetry
in occupation of left- and right chiral eigenmodes, observed through the existence of a fermion condensate. A feature of this index
theorem is that all non-trivial physics happens in the zero-mode space. In a sense, the price we pay for those modes not to
contribute to the action, is the violation of chiral symmetry. At zero as well as low temperatures,  the 't Hooft interaction is dominant, the
random instantons play an important role in chiral symmetry breaking. However, at high temperatures and near the chiral phase transition,
the instantons are no longer random, but become correlated. Therefore, it was suggested in Refs. \cite{Ilgenfritz:1988dh, Ilgenfritz:1994nt, Schafer:1994nv} that the growing correlations between instantons and anti-instantons near $T_c$ will lead to the decrease of random
instantons but the increase of instanton--anti-instanton molecule pairs. This means that the random instantons and anti-instantons
are not annihilated but paired up into the correlated $I\bar{I}$ molecules when chiral phase transition happens.
As shown in \cite{Schafer:1994nv}, in the temperature region of $T\gtrsim T_c$, the $I\bar{I}$ molecules pairing induces a repulsive
effective local four-quark interactions in the isoscalar axial-vector channel. This unconventional repulsive axial-vector interaction
leads to a repulsive axial-vector mean field in the space-like components but an attractive one in the time-like components,
which naturally induces a spontaneous local $\mathcal {CP}$ violation and local chirality imbalance as shown in \cite{Yu:2014sla}.

In Refs.  \cite{Schafer:1994nv} and \cite{Yu:2014sla}, the instanton background only couples to light u,d quarks, and the
$\mathcal{L}_{VA}$ in the isoscalar vector and axial-vector takes the form of
\beqn
{\cal L}_{VA}^{u,d}=\sum_{f=u,d}\left[G_V\left(\bar\psi_f\gamma^{\mu}\psi_f\right)^2
  +G_A \left(\bar\psi_f \gamma^{\mu} \gamma^5 \psi_f\right)^2\right].
\label{eq:L:VA-2}
\eeqn
We may generalize this term to three-flavor case as following:
\beqn
{\cal L}_{VA}^{u,d,s}=\sum_{f=u,d,s}\left[G_V\left(\bar\psi_f\gamma^{\mu}\psi_f\right)^2
  +G_A \left(\bar\psi_f \gamma^{\mu} \gamma^5 \psi_f\right)^2\right]. \nonumber \\
\label{eq:L:VA-3}
\eeqn

In reality, considering the strange quark mass is heavier than light u,d quarks, the  isoscalar vector and
axial-vector interaction induced by instanton background might take the form in between Eq.(\ref{eq:L:VA-2})
and Eq.(\ref{eq:L:VA-3}).  For example, the quark propagator in a single instanton background is associated with
the quark zero-modes \cite{'tHooft:1976fv}
\begin{equation}
\psi_0^\pm(x) = \frac{\rho}{\pi} \frac{1 \pm \gamma_5}{(r^2 +
\rho^2)^{3/2}} \, \frac{\rlap{/}{r}}{r} \, U, \label{zm},
\end{equation}
where the superscript $\pm$ corresponds to an (anti-) instanton centered at $x_0$ and with size $\rho$. The spin-color
matrix $U$ satisfies $(\vec{\sigma} + \vec{\tau}) \, U = 0$ and $r = x-x_0$. The zero mode contributions will enter
into the calculation of the correlators through the leading term in the spectral representation of the quark background
field propagator
\begin{equation}
S_{q}^\pm(x,y) = {\psi_0^\pm(x) {\psi_0^\pm}^\dagger(y)\over m_q^*
(\rho)} + O(\rho m_q^*) \; .
\label{zero}
\end{equation}
Here, the flavor dependent effective quark mass $m_q^*(\rho)=m_q-{2\over 3}
\pi^2\rho^2 \langle\overline{q}q\rangle$ (where $q$ stands for up,
down and strange quarks) in the denominator is generated by interactions with
long-wavelength QCD vacuum fields as shown in Ref. \cite{Shifman:1979uw}. It can be seen
that the strange quark propagator in the instanton background is suppressed by $1/m_s^*$.
Extending this to the instanton--anti-instanton molecule background, we can assume the actual
interaction in the isoscalar vector and axial-vector channel can be written as:
\beqn
{\cal L}_{VA}&=&\sum_{f=u,d}\left[G_V\left(\bar\psi_f\gamma^{\mu}\psi_f\right)^2
  +G_A \left(\bar\psi_f \gamma^{\mu} \gamma^5 \psi_f\right)^2\right] \nonumber \\
  & & +  \left[G_V'\left(\bar s \gamma^{\mu}s \right)^2
  +G_A' \left(\bar s \gamma^{\mu} \gamma^5 s \right)^2\right].
\label{eq:L:VA-23}
\eeqn
with $G_V'\ll G_V$ and $G_A'\ll G_A$ ($G_V'/G_V$ or $G_A'/G_A \sim (m_u/m_s)^2$ in the chirally symmetric phase by Eq.~\ref{zero}, where $m_u$ and $m_s$ represent the current masses of u and s quarks, respectively).
A compelling consequence of this argument is that the ${\cal L}_{VA}$-term, which is the essential source
of the inverse magnetic catalysis in our approach, distinguishes between the isoscalar channel with light quarks
and the one with strange quarks. It was found on the lattice QCD~\cite{Bali:2012zg} that the inverse magnetic
catalysis effect appears only for light quarks not for strange quarks, an implicit feature in our approach. In the
following calculations, we take two cases for the isoscalar vector and axial-vector interaction:
\begin{eqnarray}
& & {\rm Case~ I: ~ take}~  {\cal L}_{VA}^{u,d}\\
& & {\rm Case~ II:~ take}~ {\cal L}_{VA}^{u,d,s}.
\end{eqnarray}

Note that it has been discussed in Ref.~\cite{Yu:2014sla} that $G_S$ and $G_V$,
are expected to be positive at the whole temperature region~\cite{Klimt:1989pm, Vogl:1991qt,Klevansky:1992qe, Hatsuda:1994pi} and are assumed to be keeping
the constants fixed by the mesonic properties in QCD vacuum for simplicity, whereas
$G_A$ is expected to be positive at zero and low temperatures~\cite{Klimt:1989pm,Vogl:1991qt} and to be negative
at the temperatures above $T_c$ as a result of the interacting instanton--anti-instanton molecule model~
\cite{Ilgenfritz:1988dh, Ilgenfritz:1994nt, Schafer:1994nv}. It has been shown in Ref.~\cite{Schafer:1994nv} that a negative
$G_A$ interaction induced by the correlated instanton--anti-instanton molecule pairs near $T_c$, will give rise to
nontrivial influence on the chiral phase transition.
Therefore, we will treat $G_A$ as a free parameter in the NJL model of SU(3) flavor version
as we did in Ref.~\cite{Yu:2014sla} with two light flavors.

Working at the mean field level, one gets the thermodynamical potential per unit
volume $\Omega$ by integrating out the quark fields $\psi$ of the Lagrangian density of Eq.~(\ref{eq:L:basic}),
\begin{equation}
\Omega  =  \frac{1}{4G_S}\sum_{f=u,d,s}\sigma_f^2+\frac{K}{2G_S^3}\sigma_u\sigma_d\sigma_s-\frac{\tilde{\mu}_5^2}{4G_A} +\sum_{f=u,d,s}\Omega^f
\label{eq:MF}
 \,,
\end{equation}
where
$\sigma_f=-2G_S\langle \bar\psi_f\psi_f\rangle$ ($f=u,\,d,\,s$) and
\begin{eqnarray}
\tilde{\mu}_5 &=& -2G_A\sum_{f=u,d}\langle \bar\psi_f \gamma^0\gamma^5\psi_f\rangle, {\rm for~ Case~ I}, \nonumber \\
\tilde{\mu}_5 &=& -2G_A\sum_{f=u,d,s}\langle \bar\psi_f \gamma^0\gamma^5\psi_f\rangle, {\rm for~ Case~ II}.
\end{eqnarray}
The contributions from the fermion loop for each flavor is given by
\beqn
\Omega^f  &=&  - N_c\frac{|q_fB|}{2\pi} \sum_{s_z,k}\alpha_{s_z k}\times
\bigg[\int_{-\infty}^\infty \frac{d p_z}{2\pi} \,
  f_\Lambda^2 (p) \, \omega_{k}^f(p)\nonumber\\
  &&+2T \ln\bigl( 1+\, e^{-\omega_{k}^f/T} \bigr) \bigg]\,,\label{eq:Omega:f}
\eeqn
where
\beqn
 \omega_{k}^f = \left\{\begin{array}{ll}
 \sqrt{M_f^2 + \bigl[ |\bm p| + s_z\,\tilde{\mu}_5\, \text{sgn}(p_z) \bigr]^2} & \text{   for~~~} f=u,\,d \,\\
 \\
 \sqrt{M_f^2 + |\bm p|^2 } & \text{   for~~~} f=s, \,\\
 \end{array} \right.
\label{eq:omega}
\eeqn
are dispersion relation for the thermal eigenfrequencies with spin factors $s_z=\pm 1$ for Case I,
and
\beqn
 \omega_{k}^f = \begin{array}{ll}
 \sqrt{M_f^2 + \bigl[ |\bm p| + s_z\,\tilde{\mu}_5\, \text{sgn}(p_z) \bigr]^2} &  f=u,\,d, \, s, \\
 \end{array}
\label{eq:omega}
\eeqn
are dispersion relation for the thermal eigenfrequencies with spin factors $s_z=\pm 1$ for Case II.
The gap equations for quark mass take the following form:
\beqn
M_u & = &m_u+\sigma_u+\frac{K}{2G_S^2}\sigma_s\sigma_d \nonumber\,, \\
M_d & = &m_d+\sigma_d+\frac{K}{2G_S^2}\sigma_s\sigma_u \nonumber\,, \\
M_s & = &m_s+\sigma_s+\frac{K}{2G_S^2}\sigma_u\sigma_d \,.
\eeqn
The 3-momentum $\bm p$ in a magnetic field is given by
\beqn
\bm p^2 = p_z^2 + 2|q_f B| k\,,
\label{eq:vp}
\eeqn
and $k = 0, 1, 2, \dots $ is a non-negative integer number labeling the Landau levels.
The spin degeneracy factor is expresses as
\begin{equation}
 \alpha_{s_zk} = \left\{ \begin{array}{ll}
  \delta_{s_z,+1} & \text{   for~~~} k=0,~ qB>0, \;\\
  \delta_{s_z,-1} & \text{   for~~~} k=0,~ qB<0, \;\\
  1 & \text{   for~~~} k\neq0.
 \end{array} \right.
\end{equation}

Following Ref.~\cite{Fukushima:2010fe}  we use a smooth regularization form factor
\begin{equation}
 f_\Lambda(p) = \sqrt{\frac{\Lambda^{2N}}
  {\Lambda^{2N} + |\bm p|^{2N}}}\;,
\label{eq:f:p}
\end{equation}
where we take $N=5$. Now, by making use of Eq.~(\ref{eq:MF}),
$\sigma_f$ and $\tilde{\mu}_5$ can be determined self-consistently
as solutions to
the saddle point equations
\beqn
\frac{\partial\Omega}{\partial\sigma_f}=\frac{\partial\Omega}{\partial\tilde{\mu}_5}=0.
\label{eq:Omega:min}
\eeqn
Numerically one can obtain these solutions by a minimisation and moreover
find the true vacuum by looking at the global minimum. This will prove
essential for our model which has multiple minima breaking chiral symmetry spontaneously.

The parameters of our model, the cutoff $\Lambda$, the coupling constants $G_S$ and $K$,
and the current quark masses $m_u=m_d$ and $m_s$ are determined by fitting $f_{\pi}$,
$m_{\pi}$, $m_K$ and $m_{\eta'}$ to their empirical values by using the smooth regularization
method. We obtain $\Lambda=604.5 \text{MeV}$, $m_u=m_d=5.1 \text{MeV}$, $m_s=133.0 \text{MeV}$,
$G_S\Lambda^2=3.250$ and $K\Lambda^5=10.58$.

$G_A$ will be treated as a free parameter,
and we will perform our calculations over  a limited range of ratios
\beqn
r_A=G_A/G_S\,.
\eeqn
It is discussed in Ref.~\cite{Yu:2014sla} that when $r_A\geqslant 0$ we always obtain the ordinary magnetic catalysis effect and only when $r_A < 0$ inverse magnetic catalysis effect can been seen.

\section{Numerical Results and Discussion}

\begin{figure}
%\begin{tabular}{ccccc}
 \centerline{\includegraphics[scale=0.23]{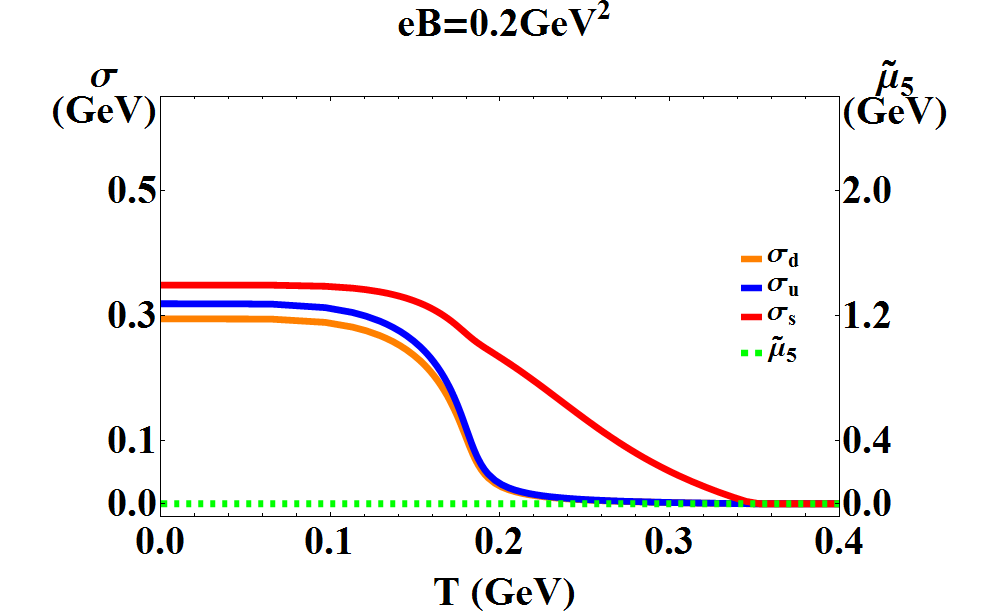}}
 \centerline{(a)
 }
\vfill
 \centerline{\includegraphics[scale=0.23]{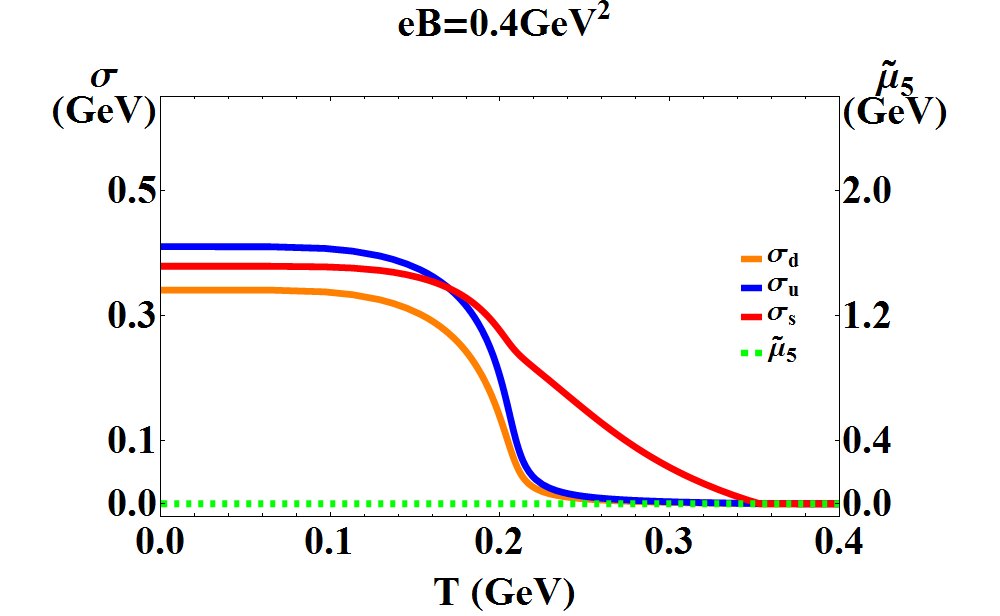}}
 \centerline{(b)
 }
 \vfill
 \centerline{\includegraphics[scale=0.23]{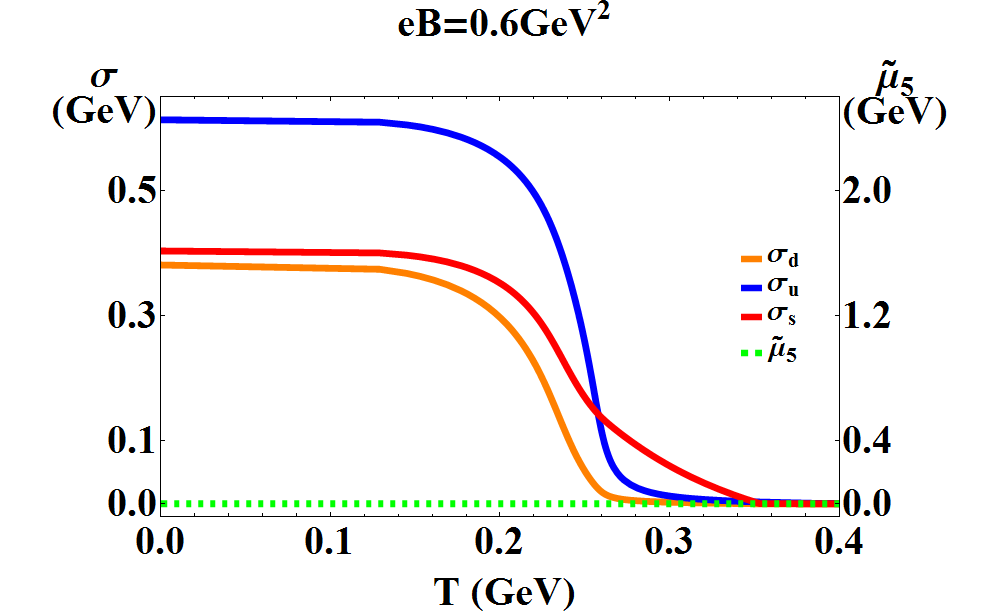}}
 \centerline{(c)
 }
%\end{tabular}
\caption{(color online). The quark condensate $\sigma_f$ (f=u,d,s)  and dynamical chiral chemical potential $\tilde\mu_5$  as a function of $T$ at $r_A=0$ for several different values of $eB$. (a) $\sigma_f$ and $\tilde{\mu}_5$ for $eB=0.2\, GeV^2$ at $r_A=0$. (b) $\sigma_f$ and $\tilde{\mu}_5$ for $eB=0.4\, GeV^2$ at $r_A=0$. (c) $\sigma_f$ and $\tilde{\mu}_5$ for $eB=0.6\, GeV^2$ at $r_A=0$.}
\label{fig:2flav:0}
\end{figure}

\begin{figure}
%\begin{tabular}{ccccc}
 \centerline{\includegraphics[scale=0.22]{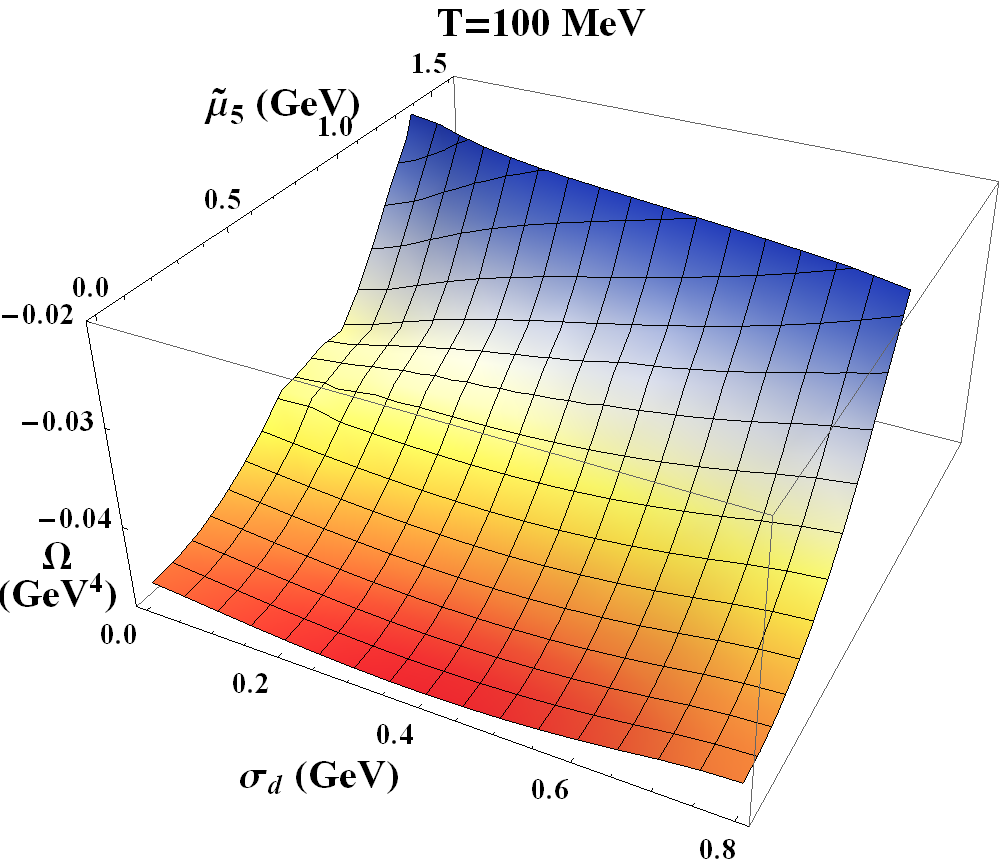}}
 \centerline{(a)
 }
\vfill
 \centerline{\includegraphics[scale=0.22]{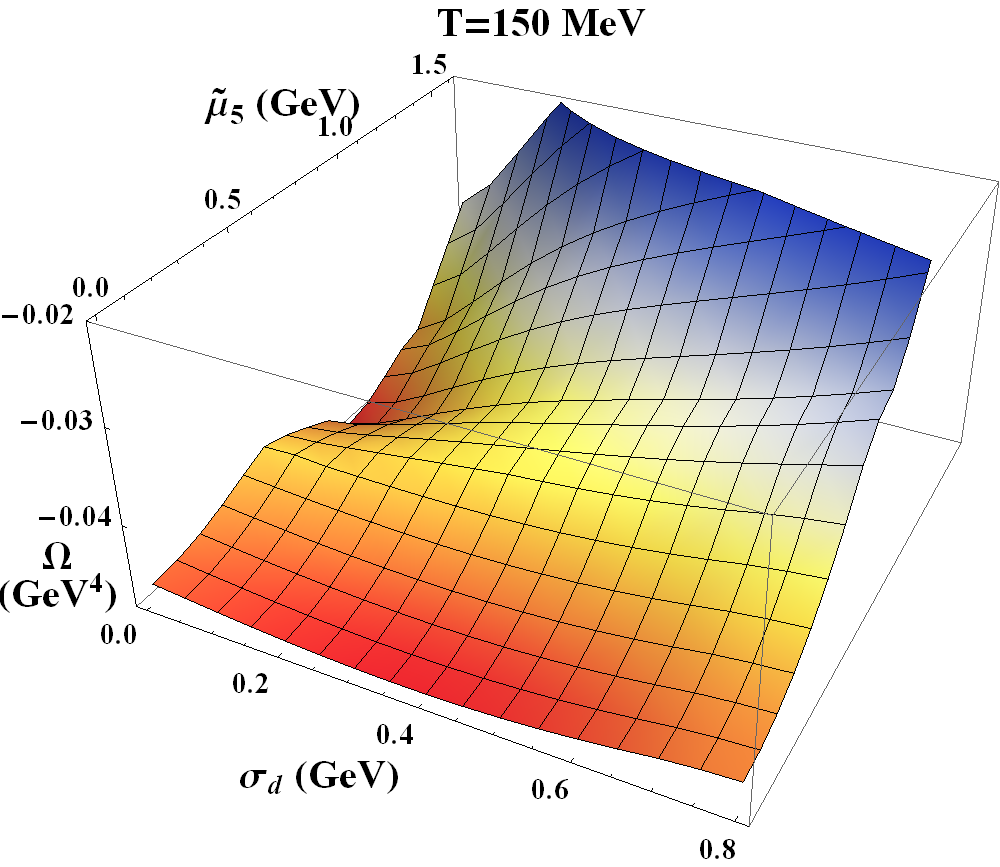}}
 \centerline{(b)
 }
 \vfill
 \centerline{\includegraphics[scale=0.22]{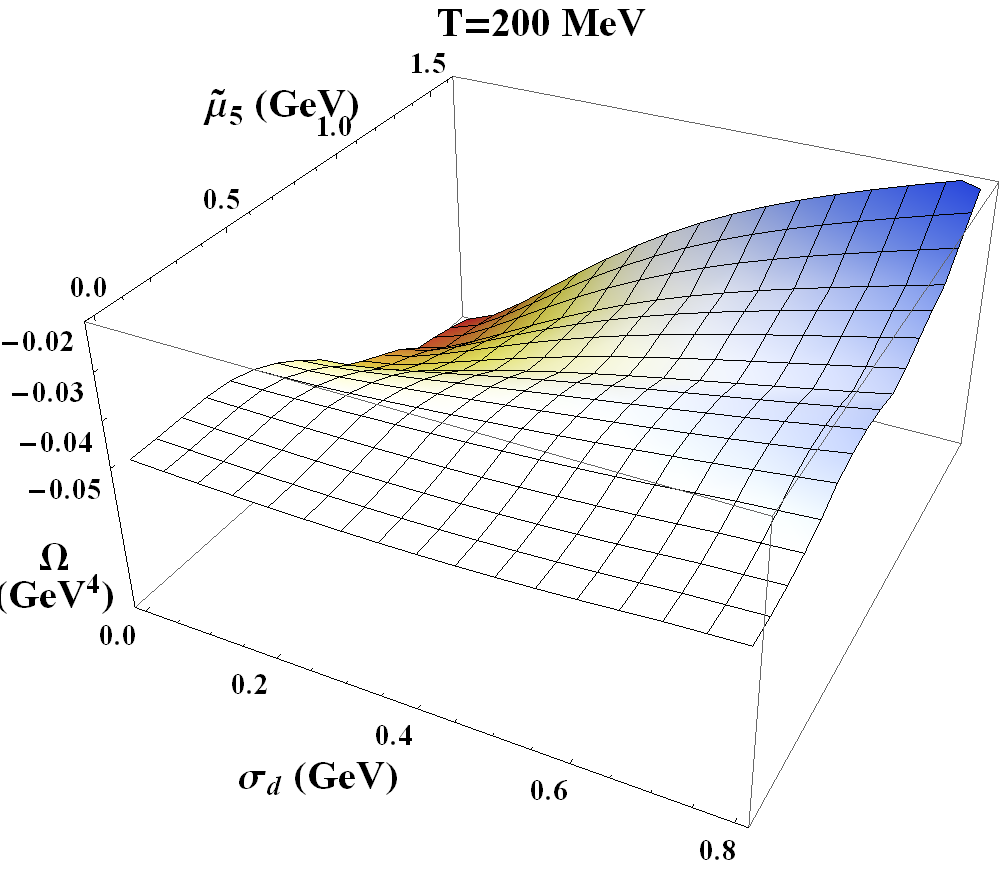}}
 \centerline{(c)
 }
%\end{tabular}
\caption{A 2D minimal surface of the potential $\Omega$ for Case I as a function of $\sigma_d$ and $\tilde\mu_5$ at $r_A=-0.5$ and $eB=0.6\,\GeV^2$ for several different values of temperature. (a) $\Omega$ at $r_A=-0.5$ and $eB=0.6\,\GeV^2$ for $T=100$ MeV (below $T_c$). (b) $\Omega$ at $r_A=-0.5$ and $eB=0.6\,\GeV^2$ for $T=150$ MeV (around $T_c$). (c) $\Omega$ at $r_A=-0.5$ and $eB=0.6\,\GeV^2$ for $T=200$ MeV (above $T_c$).}
\label{fig:surface}
\end{figure}

Considering the interaction in strange quark channel is suppressed, for our numerical calculations, we mostly take the isoscalar axial-vector
interaction of Case I, i.e. take the form in Eq.(\ref{eq:L:VA-2}).

We study the chiral phase transition at finite temperature by using Eq.~(\ref{eq:Omega:min}) for several different values of the parameters $r_A$ and $eB$, representing different intensity for the instanton--anti-instanton molecule background and the magnetic field background, respectively. Consequently, the diagrams of $\sigma-T$ for different quark flavors as well as the diagrams of $\tilde{\mu}_5-T$ can be obtained, which allow us to efficiently find the transition temperatures as functions of the parameters $r_A$ and $eB$. Actually, each of the condensates has thus its own specific transition temperature, which is defined by the temperature at the inflection point of the $\sigma-T$ diagram for each flavor, i.e., the maximum point of the quantity $\partial \sigma/\partial T$.
Here we will use $\sigma_u+\sigma_d$ to determine the transition temperature $T_c$ for the chiral phase transition of QCD.

In Fig.~\ref{fig:2flav:0}, we display the quark condensates of $\sigma_u$, $\sigma_d$ and $\sigma_s$ as well as dynamical
chiral chemical potential $\tilde{\mu}_5$ as functions of $T$ for several different values of $eB$ without considering additional axial-vector couplings, i.e., $r_A=0$ and $\tilde{\mu}_5\equiv 0$, and the ordinary magnetic catalysis effect can be exactly found from the plots. In fact, even
if we choose positive values for the paramter $r_A$, fitted by the conventional SU(3)-flavor NJL model, we will acquire the same results. This is because of the fact that when $r_A>0$, the potential energy density $\Omega$ can only has a local maximum at nonzero $\tilde{\mu}_5$, which forces $\langle\bar\psi \gamma^0\gamma^5\psi\rangle$ to be zero so that $\tilde{\mu}_5=0$ accordingly.

\begin{figure}
%\begin{tabular}{ccccc}
 \centerline{\includegraphics[scale=0.23]{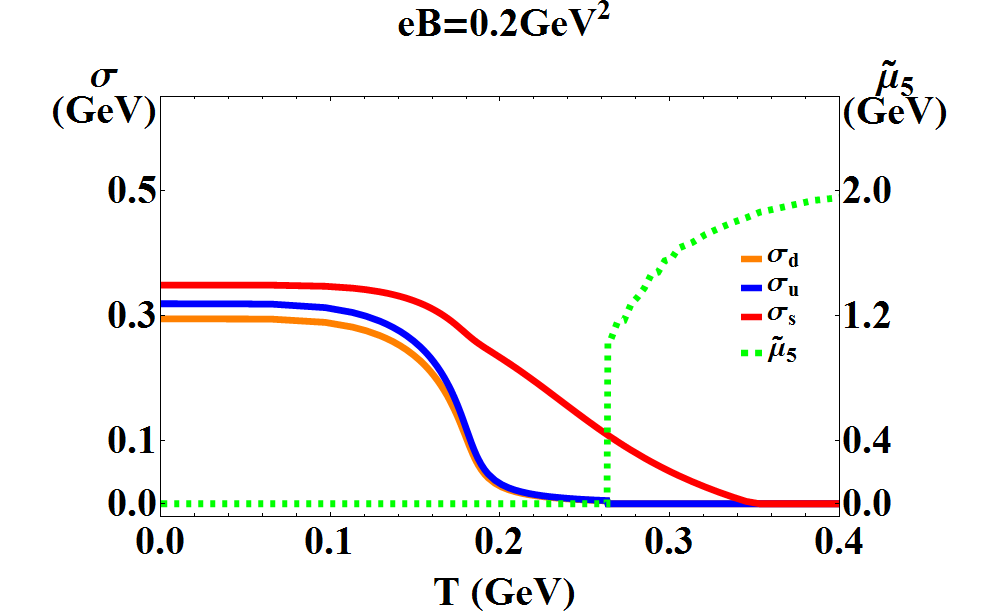}}
 \centerline{(a)
 }
\vfill
 \centerline{\includegraphics[scale=0.23]{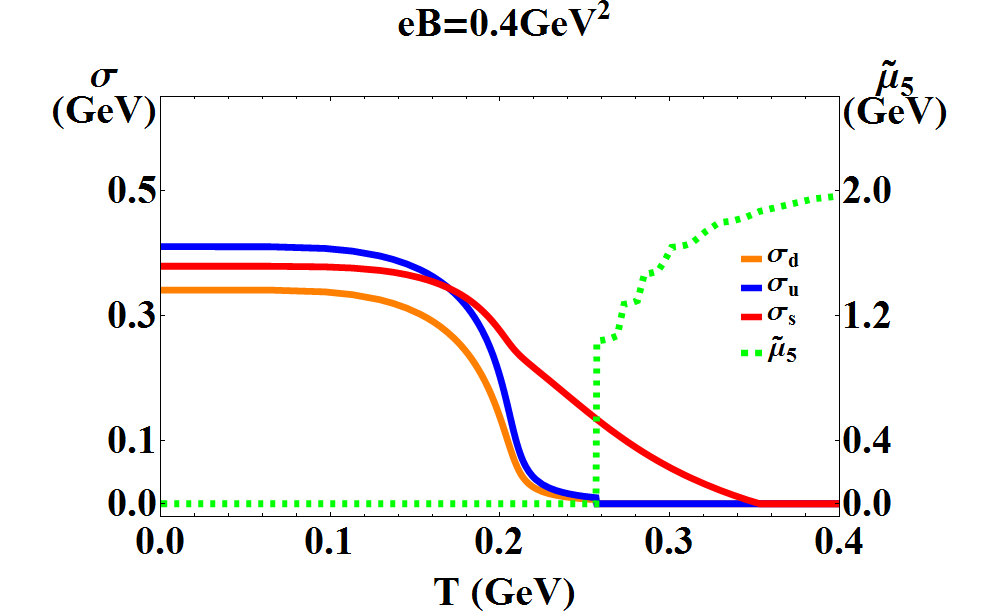}}
 \centerline{(b)
 }
 \vfill
 \centerline{\includegraphics[scale=0.23]{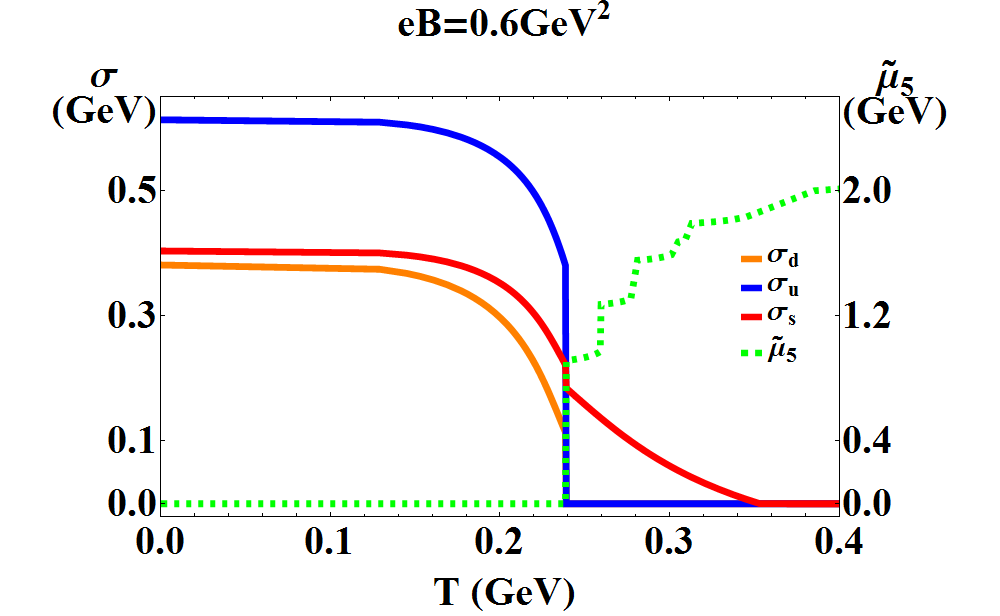}}
 \centerline{(c)
 }
%\end{tabular}
\caption{(color online). For Case I, the quark condensate $\sigma_f$ (f=u,d,s) and dynamical chiral chemical potential $\tilde\mu_5$ as a function of $T$ at $r_A=-0.3$ for several different values of $eB$. (a) $\sigma_f$ and $\tilde{\mu}_5$ for $eB=0.2\, GeV^2$ at $r_A=-0.3$. (b) $\sigma_f$ and $\tilde{\mu}_5$ for $eB=0.4\, GeV^2$ at $r_A=-0.3$. (c) $\sigma_f$ and $\tilde{\mu}_5$ for $eB=0.6\, GeV^2$ at $r_A=-0.3$.}
\label{fig:2flav:3}
\end{figure}

\begin{figure}
%\begin{tabular}{ccccc}
 \centerline{\includegraphics[scale=0.23]{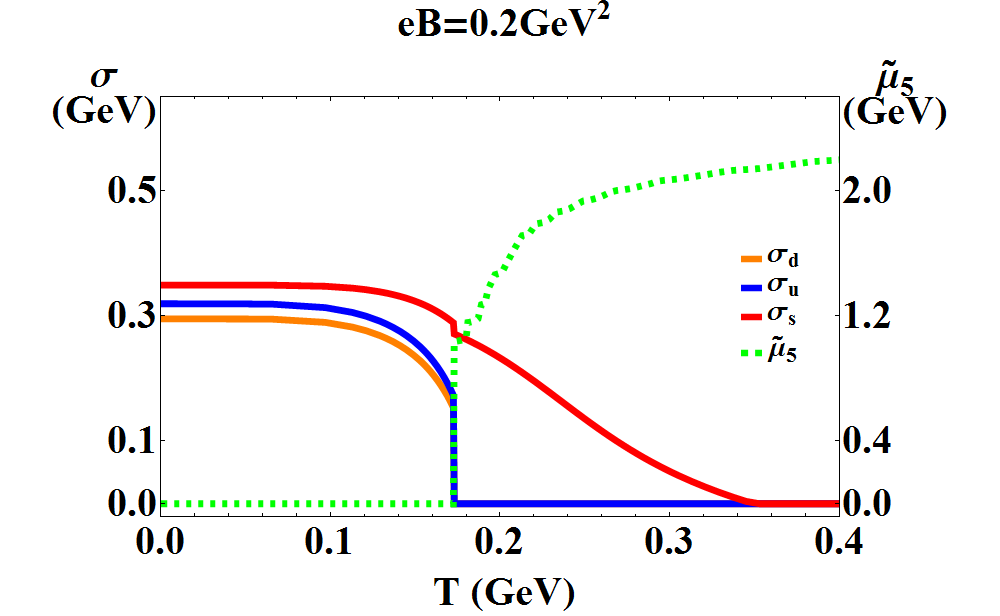}}
 \centerline{(a)
 }
\vfill
 \centerline{\includegraphics[scale=0.23]{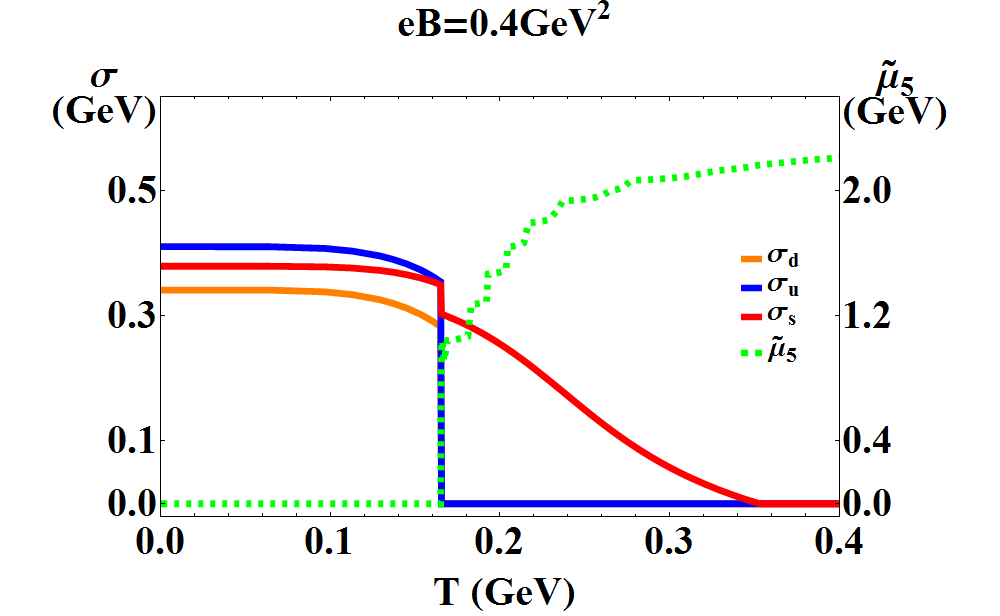}}
 \centerline{(b)
 }
 \vfill
 \centerline{\includegraphics[scale=0.23]{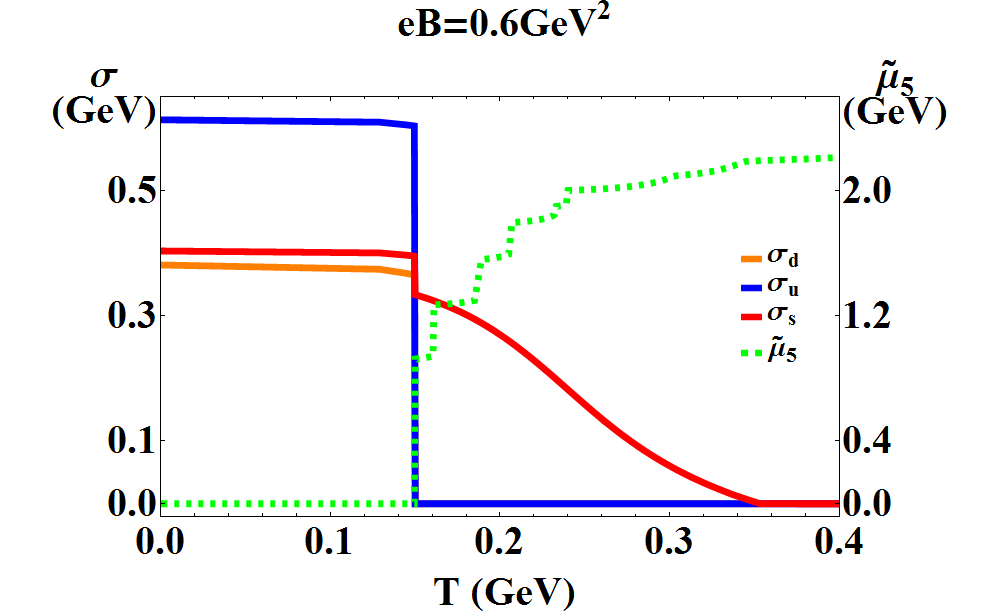}}
 \centerline{(c)
 }
%\end{tabular}
\caption{(color online). For Case I, the quark condensate $\sigma_f$ (f=u,d,s) and dynamical chiral chemical potential $\tilde\mu_5$ as a function of $T$ at $r_A=-0.5$ for several different values of $eB$. (a) $\sigma_f$ and $\tilde{\mu}_5$ for $eB=0.2\, GeV^2$ at $r_A=-0.5$. (b) $\sigma_f$ and $\tilde{\mu}_5$ for $eB=0.4\, GeV^2$ at $r_A=-0.5$. (c) $\sigma_f$ and $\tilde{\mu}_5$ for $eB=0.6\, GeV^2$ at $r_A=-0.5$.}
\label{fig:2flav:5}
\end{figure}

\begin{figure}
%\begin{tabular}{ccccc}
 \centerline{\includegraphics[scale=0.23]{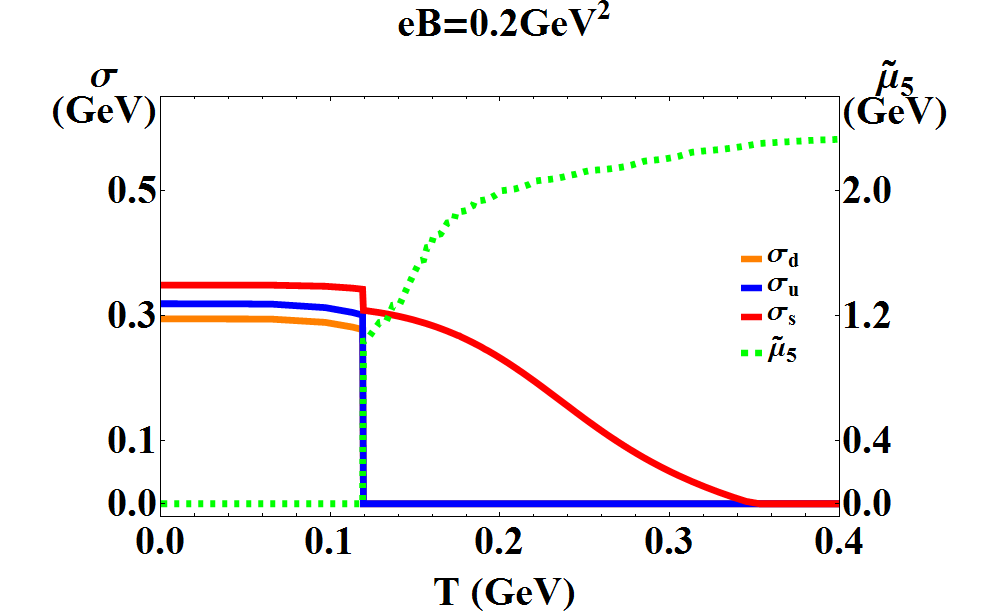}}
 \centerline{(a)
 }
\vfill
 \centerline{\includegraphics[scale=0.23]{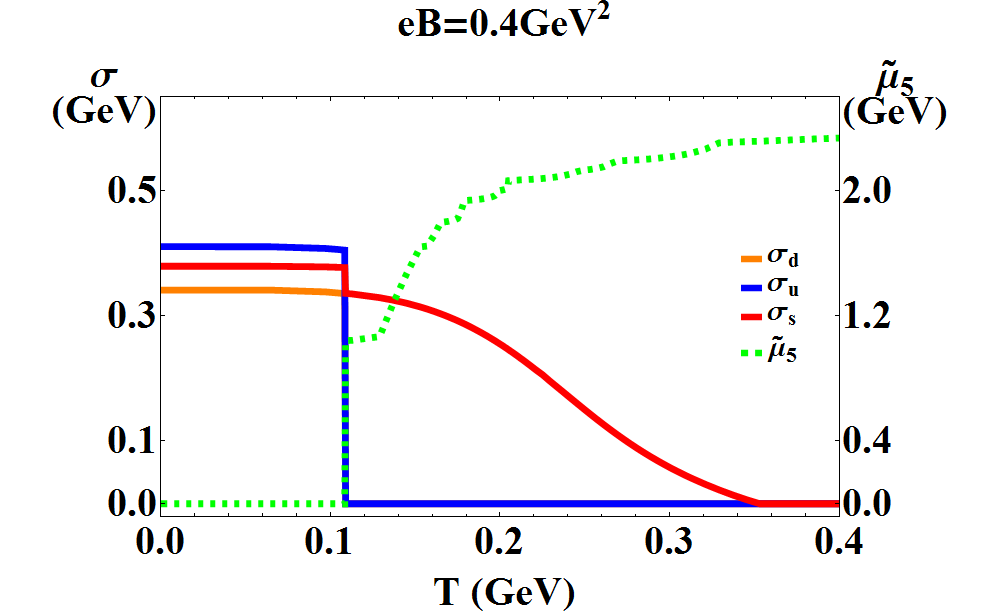}}
 \centerline{(b)
 }
 \vfill
 \centerline{\includegraphics[scale=0.23]{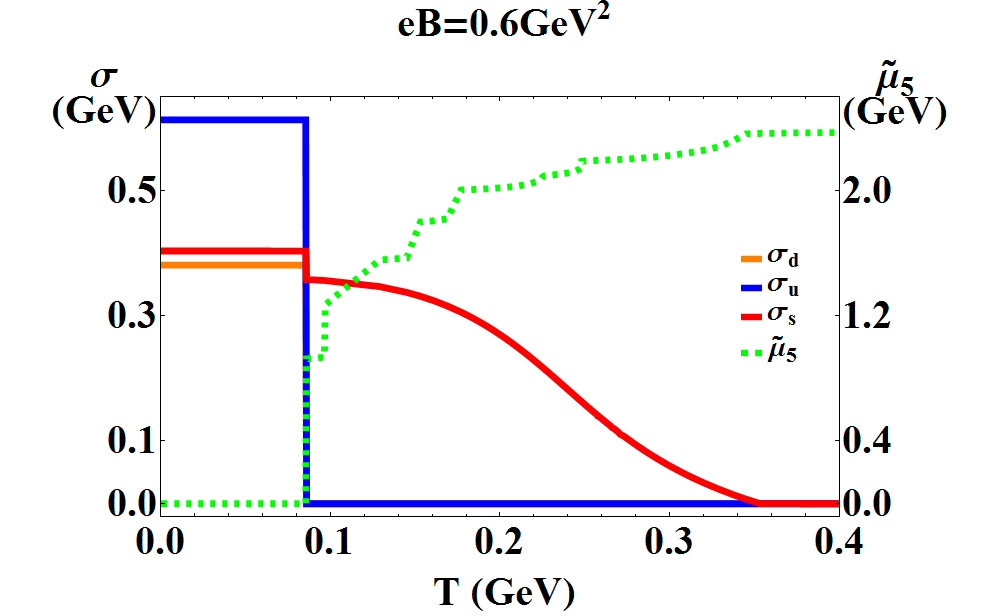}}
 \centerline{(c)
 }
%\end{tabular}
\caption{(color online). For Case I, the quark condensate $\sigma_f$ (f=u,d,s) and dynamical chiral chemical potential $\tilde\mu_5$ as a function of $T$ at $r_A=-0.7$ for several different values of $eB$. (a) $\sigma_f$ and $\tilde{\mu}_5$ for $eB=0.2\, GeV^2$ at $r_A=-0.7$. (b) $\sigma_f$ and $\tilde{\mu}_5$ for $eB=0.4\, GeV^2$ at $r_A=-0.7$. (c) $\sigma_f$ and $\tilde{\mu}_5$ for $eB=0.6\, GeV^2$ at $r_A=-0.7$.}
\label{fig:2flav:7}
\end{figure}

Next, as discussed in Ref.~\cite{Yu:2014sla}, by switching on a negative $r_A$, we introduce a non-trivial dependence of the thermodynamical potential $\Omega$ on the dynamic chiral chemical potential $\tilde\mu_5$, which adds an extra dimension to the mean field surface of $\Omega$.
As is shown in Fig.~\ref{fig:surface}, there are two local minima for the potential $\Omega$. When the temperature is low, the original local minimum representing nonzero quark condensates is dominant, since it is the global minimum. However, sufficient heating of the QCD system makes the local minimum for non-trivial $\tilde\mu_5$ energetically more favourable while the chiral condensates is weakened to the trivial state. As a result, the vacuum tunnels to a state with local chirality imbalance between right- and left-handed quarks as this metastable state becomes the new global minimum. It is important to point out that the potential $\Omega$ in Eq.~(\ref{eq:MF}) is even in $\tilde \mu_5$ so one can get separated local domains with chiral densities of both signs. Moreover, another important consequence of the competition between these two local minima is that no `mixed' state appears, so one has either nonzero quark condensates but no chirality imbalance or an instability towards the formation of nonzero dynamic chiral chemical potential but no presence of condensates. This is clear from the fact that the minima in Fig.~\ref{fig:surface} appear on either of the axes and is a consistent feature of all our simulations.

\begin{figure}
%\begin{tabular}{ccccc}
 \centerline{\includegraphics[scale=0.23]{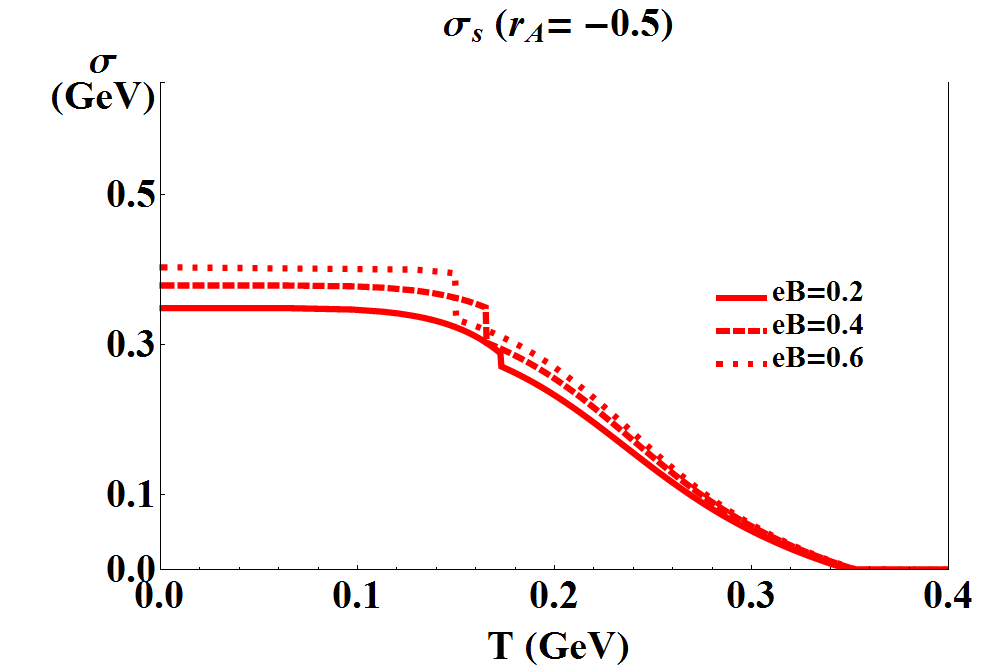}}
 \centerline{(a)
 }
\vfill
 \centerline{\includegraphics[scale=0.23]{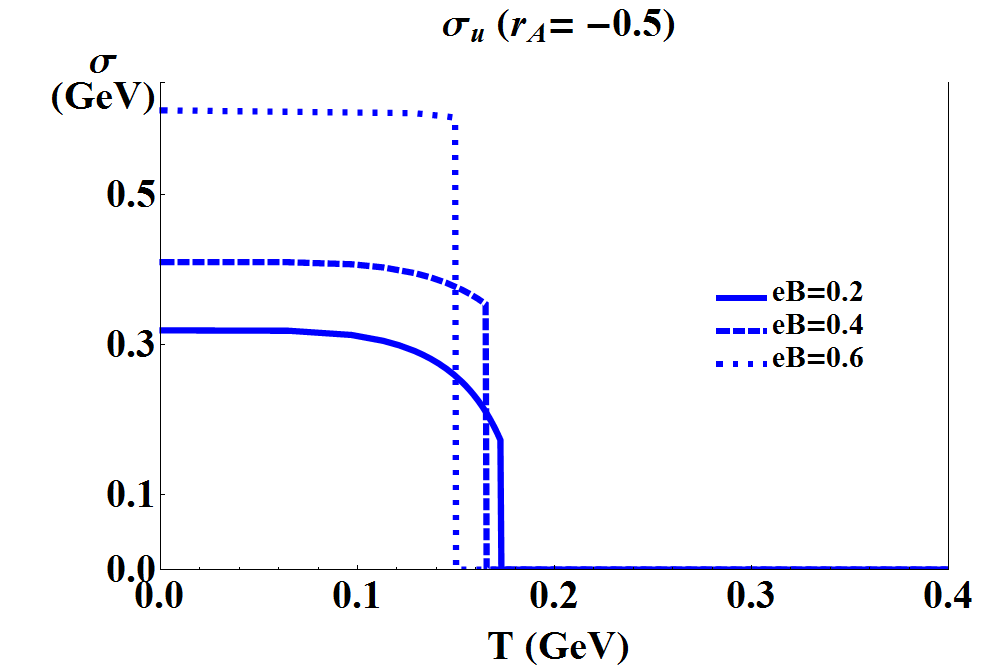}}
 \centerline{(b)
 }
 \vfill
 \centerline{\includegraphics[scale=0.23]{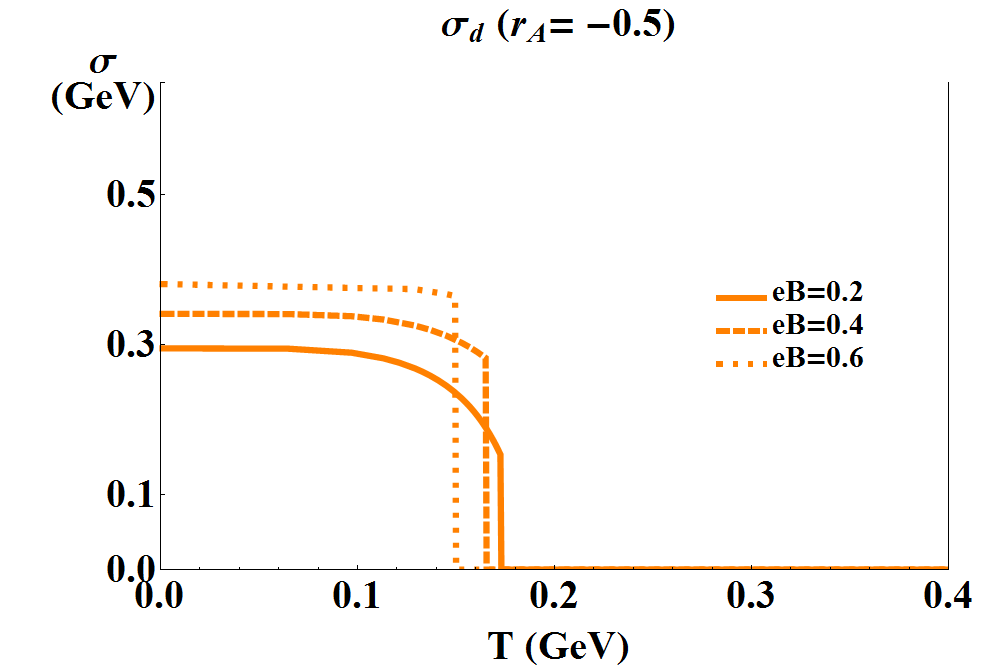}}
 \centerline{(c)
 }
%\end{tabular}
\caption{(color online). For Case I, the quark condensates $\sigma_{s}$, $\sigma_u$ and $\sigma_d$ as a function of $T$ at $r_A=-0.5$ for several different values of $eB$. (a) $\sigma_{s}$ at $r_A=-0.5$ for several different values of $eB$. (b)$\sigma_{u}$ at $r_A=-0.5$ for several different values of $eB$. (c) $\sigma_{d}$ at $r_A=-0.5$ for several different values of $eB$.}
\label{fig:2flav:recom}
\end{figure}

In fact, the magnitude of the unconventional negative $G_A$, which leads to an attractive mean field in the time-like components of the axial-vector channel, reflects the coupling strength of the attraction for $\tilde\mu_5$. In Figs.~\ref{fig:2flav:3}, \ref{fig:2flav:5} and \ref{fig:2flav:7}, we compare our numerical results of quark condensates $\sigma_f$ ($f=$ u, d, and s) and dynamic chiral chemical potential $\tilde\mu_5$ as functions of T for several magnetic fields at $r_A = -0.3$, $-0.5$ and $-0.7$. One can distinguish two distinct cases as a result of the magnitude of negative $r_A$: (i) $T_c(eB=0;\,r_A=0)< T_{5c}(eB=0)$ and (ii) $T_c(eB=0;\,r_A=0)> T_{5c}(eB=0)$.

\begin{figure}
%\begin{tabular}{ccccc}
 \centerline{\includegraphics[scale=0.23]{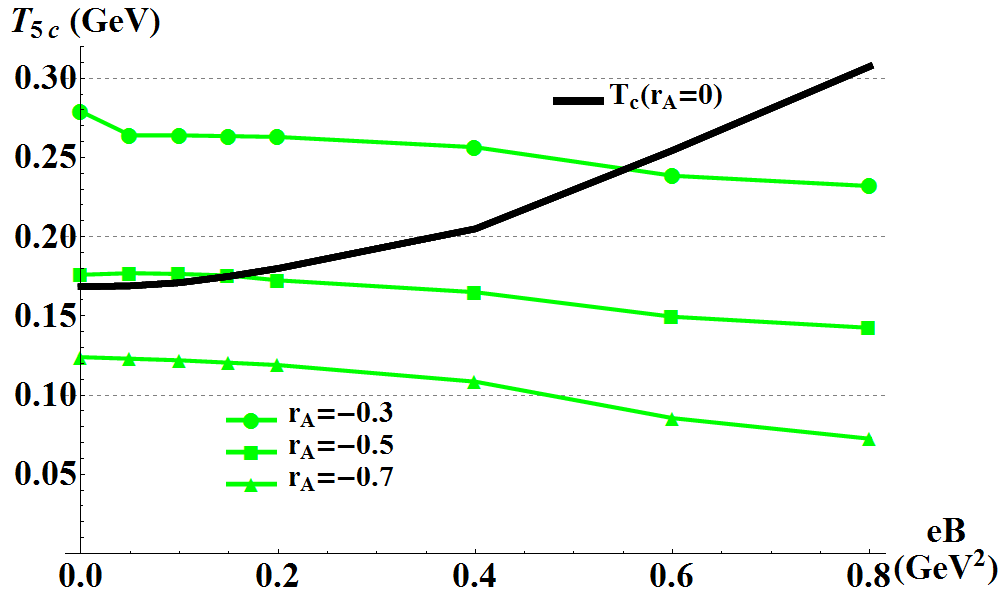}}
 \centerline{(a)
 }
\vfill
 \centerline{\includegraphics[scale=0.23]{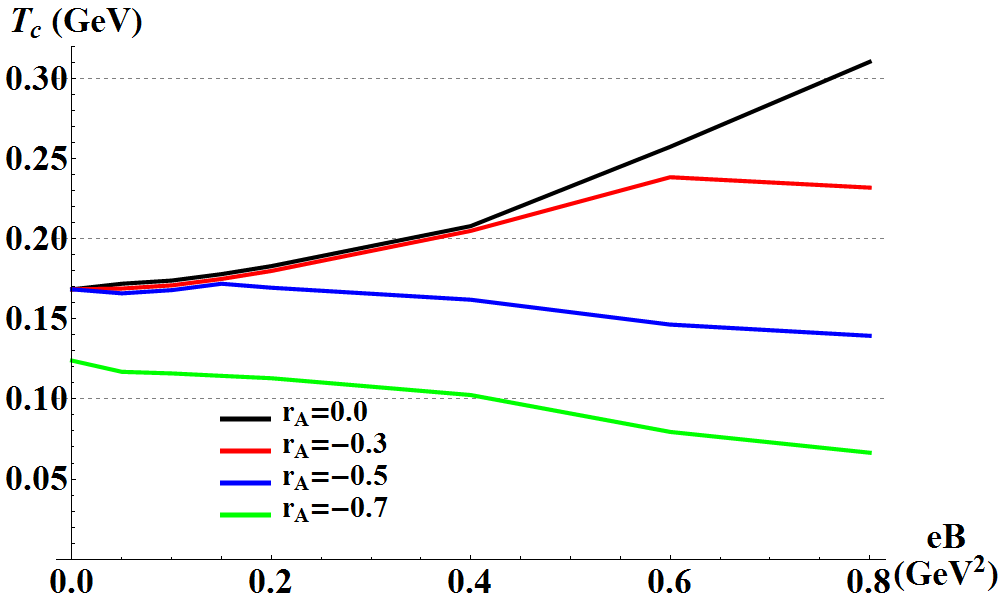}}
 \centerline{(b)
 }
 \vfill
 \centerline{\includegraphics[scale=0.23]{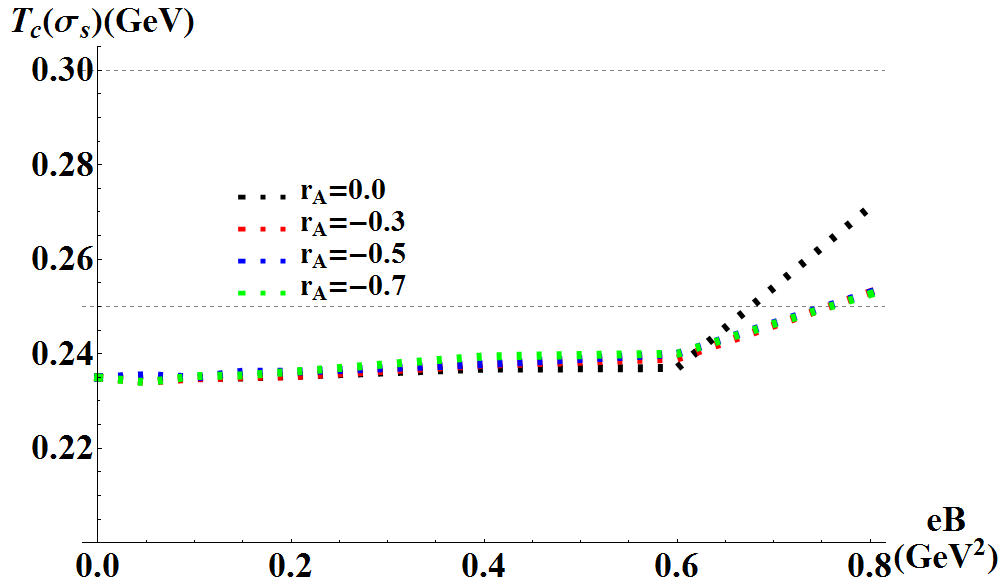}}
 \centerline{(c)
 }
%\end{tabular}
\caption{(a) The critical temperature $T_{5c}$ as a function of $eB$ for several different values of $r_A$, comparing with $T_{c}$ as a function of $eB$ at $r_A=0$.
(b) The critical temperature $T_{c}$ as a function of $eB$ for several different values of $r_A$.
(c) The critical temperature $T_{c}(\sigma_s)$ as a function of $eB$ for several different values of $r_A$. The results are for Case I.}
\label{fig:imc}
\end{figure}

If magnitude of $G_A$ is small, approximately $-0.5<r_A<0$ for our model,
it is easy to find that we are in the case (i), i.e. $T_{5c}>T_{c}(r_A=0)$ at $eB=0$. For example, as shown by Fig.~\ref{fig:2flav:3} of $r_A=-0.3$ and Fig.~\ref{fig:2flav:5} of $r_A=-0.5$, when the external magnetic field is not strong enough, the ordinary phase transition into the chirally restored phase takes place at a lower temperature and is the dominant effect in destroying the quark condensates; whereas a local $\mathcal{CP}$-odd first order phase transition for $\tilde{\mu}_5$ is spontaneously generated at a higher critical temperature $T_{5c}>T_c$.
As the magnetic field grows, both critical temperatures, $T_{5c}$ and $T_c$, approach each other and two local minima in the thermodynamic potential $\Omega$ co-exist like in the example of Fig.~\ref{fig:surface}.
At some critical value of magnetic field $B_c$ for a given $r_A$, these two critical temperatures meet with each other and the first order transition for nonzero dynamical chiral chemical potential $\tilde{\mu}_5$ becomes dominant effect, which makes $\sigma_u$ and $\sigma_d$ drop to zero at $T_c=T_{5c}$. Therefore, we find that, for u and d quarks in the case (i), the critical temperature $T_{5c}$ decreases with $eB$, while
$T_c$ increases at first and then decreases as the magnetic field grows (see Fig.~\ref{fig:imc}). As for s quark condensate, it shows a slight jump because of $\tilde{\mu}_5$ background and then continues to dissolve with increasing temperature. The critical temperature $T_c(\sigma_s)$ will increase with $eB$ always, depicted by Fig.~\ref{fig:2flav:recom}, which is consistent with the lattice results in Ref.~\cite{Bali:2012zg} in some sense.

If magnitude of $G_A$ is large enough ($r_A<-0.5$), that is in the case (ii), the light quark condensates are destroyed at $T_c=T_{5c}$ because the QCD ground state with chirality imbalanced density becomes more favorable around the critical temperature for any values of $eB$, before $\sigma_u$ and $\sigma_d$ are dissolved at their original critical temperature $T_c$ without considering $\tilde{\mu}_5$, e.g., shown by Fig.~\ref{fig:2flav:7} at $r_A=-0.7$. Hence, the critical temperatures both $T_c$ and $T_{5c}$ for u and d quark condensates decreases with $eB$ starting from $eB=0$, which is just the decreasing $T_c$ dependence on $B$ predicted by Ref.~\cite{Bali:2011qj}. On the other hand, the
condensates $\sigma_u$ and $\sigma_d$ increase with the magnetic field at zero and low temperatures still, which is the ordinary magnetic catalysis effect validated in Ref.~\cite{Bali:2012zg}. The behavior of both strange quark condensate $\sigma_s$ and the critical temperature $T_c(\sigma_s)$ is similar to that in the case (i).

We can now put the data about the critical temperatures obtained from Figs.~\ref{fig:2flav:3}, \ref{fig:2flav:5} and \ref{fig:2flav:7} into one $T_c-eB$ phase diagram of QCD. Adding more data points from identical simulations with other values of the background parameters $r_A$ and $eB$, we can find that the middle diagram of Fig.~\ref{fig:imc} shows two different possible types of dependence of $T_c$ on the magnetic field as a result of the free parameter $r_A$, which bas been discussed explicitly above. As a consequence, a reasonable strength of $r_A$, approximately between $-0.5$ and $-0.55$ by the simulations of our model, will
naturally explain the decreasing dependence of $T_c$ on $eB$ obtained in a recent lattice QCD study~\cite{Bali:2011qj}. If the magnitude of $r_A$ is too small, less than $0.5$, we can not find a monotonously decreasing dependence of $T_c$ on $eB$; If the magnitude of $r_A$ is too big, more than $0.55$, the value of $T_c$ at $eB =0$ will deviate from the lattice QCD result greatly. Furthermore, we can distinguish case (i) and case (ii) easily by the $T_{5c}(B)$ function from the top diagram of Fig.~\ref{fig:imc}, the separation given by the thick black line for the critical temperature $T_c$ at $r_A = 0$. As for the critical temperature $T_c(\sigma_s)$, depicted by the bottom diagram of Fig.~\ref{fig:imc}, one can find that its behavior at different negative values of $r_A$ is similar to that at $r_A = 0$, showing a slightly increasing dependence on $eB$.

\begin{figure}[H]
\begin{center}
\includegraphics[scale=0.22]{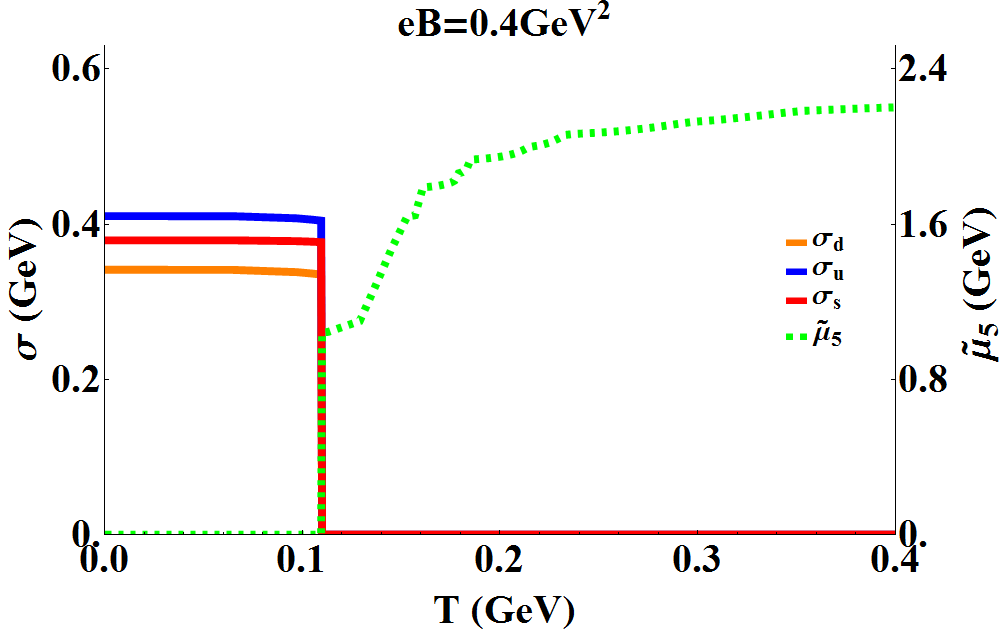}
\caption{For Case II, the quark condensate $\sigma_f$ (f=u,d,s) and dynamical chiral chemical potential $\tilde\mu_5$ as a function of $T$ at $r_A=-0.5$ and $eB=0.4\,\GeV^2$ when introducing a negative axial-vector coupling to all three flavor quarks equally.}
\label{fig:3flav:3}
\end{center}
\end{figure}

The above calculations are based on Case I, where the isoscalar axial-vector interaction only involves light u,d quarks.
The isoscalar nature of the interaction ${\cal L}_{VA}$ is essential for the nature of the phase transition. With little extra effort we were able to simulate the Case II where the four-fermion chiral attraction treats all flavors equally as given by Eq.(\ref{eq:L:VA-3}). It can be seen that the results of $\sigma_u$ and $\sigma_d$ are very similar to what we found before, but rather than a small shift in the value of the heavy strange quark condensate, $\sigma_s$ undergoes the same first order phase transition as the other two light flavors and vanishes at the transition temperature $T_{5c}$, as shown in Fig.~\ref{fig:3flav:3}. As we argued in the previous section, this kind of equal coupling with negative $G_A$ to all three quark flavors is not to be expected for an axial-vector coupling induced by an instanton--anti-instanton molecule background, and unsurprisingly it does not reproduce the qualitative lattice result.

Before drawing our final conclusions, it is important to realize we can only trust our results qualitatively. Since the new minimum and the corresponding phase transition shown in Fig.\ref{fig:surface} are in fact at a scale well beyond the cut-off of our theory, exact quantitative prediction are beyond the scope of the NJL framework. Qualitatively, however, we can be sure that the instability will emerge, and a new vacuum state will appear that is more favored than the chiral condensate when increasing magnetic fields  around $T_c$ and thus give rise to inverse magnetic catalysis effect. In that sense we think that our model is a good representation of the effect of  instanton--anti-instanton molecule background on the chiral condensates, but we cannot produce accurate predictions for the large chiral densities involved.

\section{Conclusions}

In this paper we extend our study to the QCD phase diagram as well as the behavior of quark condensates at finite temperature under an external magnetic field within the three-flavor NJL model including additional isoscalar vector and axial-vector channels. Note that an important and unconventional feature of our model is the isoscalar axial-vector interaction with a negative coupling constant depending mostly on the up and down quarks, while the interaction in the strange quark sector is suppressed due to its heavier mass, which can be derived from the instanton--anti-instanton molecule model~\cite{Schafer:1994nv}.

In this scenario, we have shown that a new way of destroying chiral condensates appears around $T_c$, replacing them by dynamical chiral chemical potential $\tilde\mu_5$ in a first order phase transition, which corresponds to a spontaneous generation of local $\mathcal {P}$ and
$\mathcal {CP}$ violation and local chirality imbalance. Moreover, the critical temperature of this first order phase transition shows inverse magnetic catalysis, meaning that it decreases with increasing magnetic field.

The dominant features of the phase transition with respect to destroying the light quark condensates depends on the parameters of the model, a tunable axial-vector coupling constant $G_A$ and the background magnetic field $B$. When increasing the magnitude of $G_A$, we can find that it will decrease the critical temperature for the first order phase transition of $\tilde\mu_5$. And the increase of the magnetic field at a given $G_A$ will also decrease the critical temperature $T_{5c}$. It means that, in the generic case, the increase of the magnitude of both parameters, $G_A$ and $eB$, will catalyze the appearance of the local chirality imbalance. Therefore, a reasonable value of $G_A$, making the ordinary chiral phase transition meet with the newly found first order phase transition at $eB=0$ (however, this is not in agreement with previous lattice results at finite temperature, and possible reasons are discussed in Ref.~\cite{Yu:2014sla}), the inverse magnetic catalysis effect can be naturally explained and a phase diagram is reproduced in Fig.~\ref{fig:imc}, consistent with lattice QCD results~\cite{Bali:2011qj}.

On the other hand, since the lattice results of Ref.~\cite{Bali:2012zg} indicated that strange quark condensate experience magnetic catalysis only, we investigate the behavior of strange quark condensate in our model also. When we introduce
a negative axial-vector interaction channel including light-quark currents only, it is found that the critical temperature $T_c(\sigma_s)$ exhibits little modification as a result of the instanton--anti-instanton molecule background, in some sense consistent with lattice results, although the strange condensate shows a slight jump arising from the appearance of $\tilde\mu_5$ in the $\sigma-T$ diagrams. This might be improved by considering the spatial structure of the topological density distribution, which is one of our future plans. In reality, the interaction in the isoscalar vector and axial-vector channel might look like Eq.(\ref{eq:L:VA-23}), with a dominant isoscalar axial-vector interaction in the light quark sector and with a suppressed interaction in the strange quark sector. In this case, if the interaction in the strange sector is small enough, the result of chiral phase transitions will be the same as that in Case I,
and we can get the inverse magnetic catalysis for light quark condensate but magnetic catalysis for strange quark condensate around critical temperature, which is in agreement with lattice results.

\vskip 0.2cm
\section*{Acknowledgement}

We thank valuable discussions with M. Chernodub, J.Y. Chao and I. Shovkovy.
This work is supported by the NSFC under Grant No. 11275213, and 11261130311(CRC 110 by DFG and NSFC),
 CAS key project KJCX2-EW-N01, and Youth Innovation Promotion Association of CAS.
L.Yu is partially supported by China Postdoctoral Science Foundation under
Grant No. 2014M550841.
The work of JVD was supported by a grant from La Region Centre (France) and the Chinese-French Cai Yuanpei 2013 grant.


\begin{thebibliography}{99}

%\cite{Skokov:2009qp}
\bibitem{Skokov:2009qp}
  V.~Skokov, A.~Y.~.Illarionov and V.~Toneev,
  %``Estimate of the magnetic field strength in heavy-ion collisions,''
  Int.\ J.\ Mod.\ Phys.\ A {\bf 24}, 5925 (2009)
  [arXiv:0907.1396 [nucl-th]].
  %%CITATION = ARXIV:0907.1396;%%
  %228 citations counted in INSPIRE as of 08 Apr 2014

%\cite{Voronyuk:2011jd}
\bibitem{Voronyuk:2011jd}
  V.~Voronyuk, V.~D.~Toneev, W.~Cassing, E.~L.~Bratkovskaya, V.~P.~Konchakovski and S.~A.~Voloshin,
  %``(Electro-)Magnetic field evolution in relativistic heavy-ion collisions,''
  Phys.\ Rev.\ C {\bf 83}, 054911 (2011)
  [arXiv:1103.4239 [nucl-th]].
  %%CITATION = ARXIV:1103.4239;%%
  %68 citations counted in INSPIRE as of 08 Apr 2014


%\cite{Bzdak:2011yy}
\bibitem{Bzdak:2011yy}
  A.~Bzdak and V.~Skokov,
  %``Event-by-event fluctuations of magnetic and electric fields in heavy ion collisions,''
  Phys.\ Lett.\ B {\bf 710}, 171 (2012)
  [arXiv:1111.1949 [hep-ph]].
  %%CITATION = ARXIV:1111.1949;%%
  %71 citations counted in INSPIRE as of 08 Apr 2014

%\cite{Deng:2012pc}
\bibitem{Deng:2012pc}
  W.~-T.~Deng and X.~-G.~Huang,
  %``Event-by-event generation of electromagnetic fields in heavy-ion collisions,''
  Phys.\ Rev.\ C {\bf 85}, 044907 (2012)
  [arXiv:1201.5108 [nucl-th]].
  %%CITATION = ARXIV:1201.5108;%%
  %64 citations counted in INSPIRE as of 08 Apr 2014

%\cite{Vachaspati:1991nm}
\bibitem{Vachaspati:1991nm}
  T.~Vachaspati,
  %``Magnetic fields from cosmological phase transitions,''
  Phys.\ Lett.\ B {\bf 265}, 258 (1991).
  %%CITATION = PHLTA,B265,258;%%
  %334 citations counted in INSPIRE as of 27 Oct 2014

%\cite{Enqvist:1993np}
\bibitem{Enqvist:1993np}
  K.~Enqvist and P.~Olesen,
  %``On primordial magnetic fields of electroweak origin,''
  Phys.\ Lett.\ B {\bf 319}, 178 (1993)
  [hep-ph/9308270].
  %%CITATION = HEP-PH/9308270;%%
  %97 citations counted in INSPIRE as of 27 Oct 2014


%\cite{Duncan:1992hi}
\bibitem{Duncan:1992hi}
  R.~C.~Duncan and C.~Thompson,
  %``Formation of very strongly magnetized neutron stars - implications for gamma-ray bursts,''
  Astrophys.\ J.\  {\bf 392}, L9 (1992).
  %%CITATION = ASJOA,392,L9;%%
  %965 citations counted in INSPIRE as of 27 Oct 2014

%\cite{Klevansky:1989vi}
\bibitem{Klevansky:1989vi}
  S.~P.~Klevansky and R.~H.~Lemmer,
  %``Chiral symmetry restoration in the Nambu-Jona-Lasinio model with a constant electromagnetic field,''
  Phys.\ Rev.\ D {\bf 39}, 3478 (1989).
  %%CITATION = PHRVA,D39,3478;%%
  %154 citations counted in INSPIRE as of 08 Apr 2014


%\cite{Klimenko:1990rh}
\bibitem{Klimenko:1990rh}
  K.~G.~Klimenko,
  %``Three-dimensional Gross-Neveu model in an external magnetic field,''
  Theor.\ Math.\ Phys.\  {\bf 89}, 1161 (1992)
  [Teor.\ Mat.\ Fiz.\  {\bf 89}, 211 (1991)].
  %%CITATION = TMPHA,89,1161;%%
  %124 citations counted in INSPIRE as of 08 Apr 2014

%\cite{Gusynin:1995nb}
\bibitem{Gusynin:1995nb}
  V.~P.~Gusynin, V.~A.~Miransky and I.~A.~Shovkovy,
  %``Dimensional reduction and catalysis of dynamical symmetry breaking by a magnetic field,''
  Nucl.\ Phys.\ B {\bf 462}, 249 (1996)
  [hep-ph/9509320].
  %%CITATION = HEP-PH/9509320;%%
  %251 citations counted in INSPIRE as of 08 Apr 2014

%\cite{Shovkovy:2012zn}
\bibitem{Shovkovy:2012zn}
  I.~A.~Shovkovy,
  %``Magnetic Catalysis: A Review,''
  Lect.\ Notes Phys.\  {\bf 871}, 13 (2013)
  [arXiv:1207.5081 [hep-ph]].
  %%CITATION = ARXIV:1207.5081;%%
  %54 citations counted in INSPIRE as of 26 Aug 2014





%\cite{Shushpanov:1997sf}
\bibitem{Shushpanov:1997sf}
  I.~A.~Shushpanov and A.~V.~Smilga,
  %``Quark condensate in a magnetic field,''
  Phys.\ Lett.\ B {\bf 402}, 351 (1997)
  [hep-ph/9703201].
  %%CITATION = HEP-PH/9703201;%%
  %142 citations counted in INSPIRE as of 08 Apr 2014



%\cite{Agasian:1999sx}
\bibitem{Agasian:1999sx}
  N.~O.~Agasian and I.~A.~Shushpanov,
  %``The Quark and gluon condensates and low-energy QCD theorems in a magnetic field,''
  Phys.\ Lett.\ B {\bf 472}, 143 (2000)
  [hep-ph/9911254].
  %%CITATION = HEP-PH/9911254;%%
  %54 citations counted in INSPIRE as of 08 Apr 2014

%\cite{Alexandre:2000yf}
\bibitem{Alexandre:2000yf}
  J.~Alexandre, K.~Farakos and G.~Koutsoumbas,
  %``Magnetic catalysis in QED(3) at finite temperature: Beyond the constant mass approximation,''
  Phys.\ Rev.\ D {\bf 63}, 065015 (2001)
  [hep-th/0010211].
  %%CITATION = HEP-TH/0010211;%%
  %21 citations counted in INSPIRE as of 08 Apr 2014



%\cite{Agasian:2001hv}
\bibitem{Agasian:2001hv}
  N.~O.~Agasian,
  %``Chiral thermodynamics in a magnetic field,''
  Phys.\ Atom.\ Nucl.\  {\bf 64}, 554 (2001)
  [Yad.\ Fiz.\  {\bf 64}, 608 (2001)]
  [hep-ph/0112341].
  %%CITATION = HEP-PH/0112341;%%
  %28 citations counted in INSPIRE as of 08 Apr 2014





%\cite{Cohen:2007bt}
\bibitem{Cohen:2007bt}
  T.~D.~Cohen, D.~A.~McGady and E.~S.~Werbos,
  %``The Chiral condensate in a constant electromagnetic field,''
  Phys.\ Rev.\ C {\bf 76}, 055201 (2007)
  [arXiv:0706.3208 [hep-ph]].
  %%CITATION = ARXIV:0706.3208;%%
  %76 citations counted in INSPIRE as of 08 Apr 2014

%\cite{Gatto:2010qs}
\bibitem{Gatto:2010qs}
  R.~Gatto and M.~Ruggieri,
  %``Dressed Polyakov loop and phase diagram of hot quark matter under magnetic field,''
  Phys.\ Rev.\ D {\bf 82}, 054027 (2010)
  [arXiv:1007.0790 [hep-ph]].
  %%CITATION = ARXIV:1007.0790;%%
  %90 citations counted in INSPIRE as of 08 Apr 2014

%\cite{Gatto:2010pt}
\bibitem{Gatto:2010pt}
  R.~Gatto and M.~Ruggieri,
  %``Deconfinement and Chiral Symmetry Restoration in a Strong Magnetic Background,''
  Phys.\ Rev.\ D {\bf 83}, 034016 (2011)
  [arXiv:1012.1291 [hep-ph]].
  %%CITATION = ARXIV:1012.1291;%%
  %107 citations counted in INSPIRE as of 08 Apr 2014

%\cite{Mizher:2010zb}
\bibitem{Mizher:2010zb}
  A.~J.~Mizher, M.~N.~Chernodub and E.~S.~Fraga,
  %``Phase diagram of hot QCD in an external magnetic field: possible splitting of deconfinement and chiral transitions,''
  Phys.\ Rev.\ D {\bf 82}, 105016 (2010)
  [arXiv:1004.2712 [hep-ph]].
  %%CITATION = ARXIV:1004.2712;%%
  %136 citations counted in INSPIRE as of 08 Apr 2014


%\cite{Kashiwa:2011js}
\bibitem{Kashiwa:2011js}
  K.~Kashiwa,
  %``Entanglement between chiral and deconfinement transitions under strong uniform magnetic background field,''
  Phys.\ Rev.\ D {\bf 83}, 117901 (2011)
  [arXiv:1104.5167 [hep-ph]].
  %%CITATION = ARXIV:1104.5167;%%
  %35 citations counted in INSPIRE as of 08 Apr 2014


%\cite{Avancini:2012ee}
\bibitem{Avancini:2012ee}
  S.~S.~Avancini, D.~P.~Menezes, M.~B.~Pinto and C.~Providencia,
  %``The QCD Critical End Point Under Strong Magnetic Fields,''
  Phys.\ Rev.\ D {\bf 85}, 091901 (2012)
  [arXiv:1202.5641 [hep-ph]].
  %%CITATION = ARXIV:1202.5641;%%
  %25 citations counted in INSPIRE as of 08 Apr 2014


%\cite{Andersen:2012dz}
\bibitem{Andersen:2012dz}
  J.~O.~Andersen,
  %``Thermal pions in a magnetic background,''
  Phys.\ Rev.\ D {\bf 86}, 025020 (2012)
  [arXiv:1202.2051 [hep-ph]].
  %%CITATION = ARXIV:1202.2051;%%
  %23 citations counted in INSPIRE as of 08 Apr 2014


%\cite{Scherer:2012nn}
\bibitem{Scherer:2012nn}
  D.~D.~Scherer and H.~Gies,
  %``Renormalization Group Study of Magnetic Catalysis in the 3d Gross-Neveu Model,''
  Phys.\ Rev.\ B {\bf 85}, 195417 (2012)
  [arXiv:1201.3746 [cond-mat.str-el]].
  %%CITATION = ARXIV:1201.3746;%%
  %12 citations counted in INSPIRE as of 08 Apr 2014



%\cite{Buividovich:2008wf}
\bibitem{Buividovich:2008wf}
  P.~V.~Buividovich, M.~N.~Chernodub, E.~V.~Luschevskaya and M.~I.~Polikarpov,
  %``Numerical study of chiral symmetry breaking in non-Abelian gauge theory with background magnetic field,''
  Phys.\ Lett.\ B {\bf 682}, 484 (2010)
  [arXiv:0812.1740 [hep-lat]].
  %%CITATION = ARXIV:0812.1740;%%
  %71 citations counted in INSPIRE as of 09 Apr 2014


%\cite{Braguta:2010ej}
\bibitem{Braguta:2010ej}
  V.~V.~Braguta, P.~V.~Buividovich, T.~Kalaydzhyan, S.~V.~Kuznetsov and M.~I.~Polikarpov,
  %``The Chiral Magnetic Effect and chiral symmetry breaking in SU(3) quenched lattice gauge theory,''
  Phys.\ Atom.\ Nucl.\  {\bf 75}, 488 (2012)
  [arXiv:1011.3795 [hep-lat]].
  %%CITATION = ARXIV:1011.3795;%%
  %40 citations counted in INSPIRE as of 09 Apr 2014


%\cite{D'Elia:2010nq}
\bibitem{D'Elia:2010nq}
  M.~D'Elia, S.~Mukherjee and F.~Sanfilippo,
  %``QCD Phase Transition in a Strong Magnetic Background,''
  Phys.\ Rev.\ D {\bf 82}, 051501 (2010)
  [arXiv:1005.5365 [hep-lat]].
  %%CITATION = ARXIV:1005.5365;%%
  %127 citations counted in INSPIRE as of 09 Apr 2014

%\cite{D'Elia:2011zu}
\bibitem{D'Elia:2011zu}
  M.~D'Elia and F.~Negro,
  %``Chiral Properties of Strong Interactions in a Magnetic Background,''
  Phys.\ Rev.\ D {\bf 83}, 114028 (2011)
  [arXiv:1103.2080 [hep-lat]].
  %%CITATION = ARXIV:1103.2080;%%
  %75 citations counted in INSPIRE as of 09 Apr 2014


%\cite{Ilgenfritz:2012fw}
\bibitem{Ilgenfritz:2012fw}
  E.~-M.~Ilgenfritz, M.~Kalinowski, M.~Muller-Preussker, B.~Petersson and A.~Schreiber,
  %``Two-color QCD with staggered fermions at finite temperature under the influence of a magnetic field,''
  Phys.\ Rev.\ D {\bf 85}, 114504 (2012)
  [arXiv:1203.3360 [hep-lat]].
  %%CITATION = ARXIV:1203.3360;%%
  %36 citations counted in INSPIRE as of 09 Apr 2014



%\cite{Bali:2011qj}
\bibitem{Bali:2011qj}
  G.~S.~Bali, F.~Bruckmann, G.~Endrodi, Z.~Fodor, S.~D.~Katz, S.~Krieg, A.~Schafer and K.~K.~Szabo,
  %``The QCD phase diagram for external magnetic fields,''
  JHEP {\bf 1202}, 044 (2012)
  [arXiv:1111.4956 [hep-lat]].
  %%CITATION = ARXIV:1111.4956;%%
  %123 citations counted in INSPIRE as of 09 Apr 2014


%\cite{Bali:2012zg}
\bibitem{Bali:2012zg}
  G.~S.~Bali, F.~Bruckmann, G.~Endrodi, Z.~Fodor, S.~D.~Katz and A.~Schafer,
  %``QCD quark condensate in external magnetic fields,''
  Phys.\ Rev.\ D {\bf 86}, 071502 (2012)
  [arXiv:1206.4205 [hep-lat]].
  %%CITATION = ARXIV:1206.4205;%%
  %83 citations counted in INSPIRE as of 09 Apr 2014




%\cite{Fukushima:2012kc}
\bibitem{Fukushima:2012kc}
  K.~Fukushima and Y.~Hidaka,
  %``Magnetic Catalysis vs Magnetic Inhibition,''
  Phys.\ Rev.\ Lett.\  {\bf 110}, 031601 (2013)
  [arXiv:1209.1319 [hep-ph]].
  %%CITATION = ARXIV:1209.1319;%%
  %30 citations counted in INSPIRE as of 09 Apr 2014


%\cite{Kojo:2012js}
\bibitem{Kojo:2012js}
  T.~Kojo and N.~Su,
  %``The quark mass gap in a magnetic field,''
  Phys.\ Lett.\ B {\bf 720}, 192 (2013)
  [arXiv:1211.7318 [hep-ph]].
  %%CITATION = ARXIV:1211.7318;%%
  %18 citations counted in INSPIRE as of 09 Apr 2014


%\cite{Bruckmann:2013oba}
\bibitem{Bruckmann:2013oba}
  F.~Bruckmann, G.~Endrodi and T.~G.~Kovacs,
  %``Inverse magnetic catalysis and the Polyakov loop,''
  JHEP {\bf 1304}, 112 (2013)
  [arXiv:1303.3972 [hep-lat]].
  %%CITATION = ARXIV:1303.3972;%%
  %34 citations counted in INSPIRE as of 09 Apr 2014

%\cite{Chao:2013qpa}
\bibitem{Chao:2013qpa}
  J.~Chao, P.~Chu and M.~Huang,
  %``Inverse magnetic catalysis induced by sphalerons,''
  Phys.\ Rev.\ D {\bf 88}, 054009 (2013)
  [arXiv:1305.1100 [hep-ph]].
  %%CITATION = ARXIV:1305.1100;%%
  %10 citations counted in INSPIRE as of 09 Apr 2014

%\cite{Fraga:2013ova}
\bibitem{Fraga:2013ova}
  E.~S.~Fraga, B.~W.~Mintz and J.~Schaffner-Bielich,
  %``A search for inverse magnetic catalysis in thermal quark-meson models,''
  Phys.\ Lett.\ B {\bf 731}, 154 (2014)
  [arXiv:1311.3964 [hep-ph]].
  %%CITATION = ARXIV:1311.3964;%%
  %4 citations counted in INSPIRE as of 23 Jun 2014

%\cite{Ferreira:2014kpa}
\bibitem{Ferreira:2014kpa}
  M.~Ferreira, P.~Costa, O.~Louren\c{c}o, T.~Frederico and C.~Provid\^{e}ncia,
  %``Inverse magnetic catalysis in the (2+1)-flavor Nambu--Jona-Lasinio and Polyakov--Nambu--Jona-Lasinio models,''
  Phys.\ Rev.\ D {\bf 89}, 116011 (2014)
  [arXiv:1404.5577 [hep-ph]].
  %%CITATION = ARXIV:1404.5577;%%
  %16 citations counted in INSPIRE as of 27 Oct 2014

%\cite{Farias:2014eca}
\bibitem{Farias:2014eca}
  R.~L.~S.~Farias, K.~P.~Gomes, G.~I.~Krein and M.~B.~Pinto,
  %``Importance of asymptotic freedom for the pseudocritical temperature in magnetized quark matter,''
  Phys.\ Rev.\ C {\bf 90}, no. 2, 025203 (2014)
  [arXiv:1404.3931 [hep-ph]].
  %%CITATION = ARXIV:1404.3931;%%
  %17 citations counted in INSPIRE as of 10 Dec 2014
  
  
%\cite{Yu:2014sla}
\bibitem{Yu:2014sla}
  L.~Yu, H.~Liu and M.~Huang,
  %``Spontaneous generation of local CP violation and inverse magnetic catalysis,''
  Phys.\ Rev.\ D {\bf 90}, 074009 (2014)
  [arXiv:1404.6969 [hep-ph]].
  %%CITATION = ARXIV:1404.6969;%%
  %7 citations counted in INSPIRE as of 28 Oct 2014

%\cite{Andersen:2014oaa}
\bibitem{Andersen:2014oaa}
  J.~O.~Andersen, W.~R.~Naylor and A.~Tranberg,
  %``Inverse magnetic catalysis and regularization in the quark-meson model,''
  arXiv:1410.5247 [hep-ph].
  %%CITATION = ARXIV:1410.5247;%%
  %3 citations counted in INSPIRE as of 10 Dec 2014

%\cite{Ferrer:2014qka}
\bibitem{Ferrer:2014qka}
  E.~J.~Ferrer, V.~de la Incera and X.~J.~Wen,
  %``Quark Antiscreening at Strong Magnetic Field and Inverse Magnetic Catalysis,''
  arXiv:1407.3503 [nucl-th].
  %%CITATION = ARXIV:1407.3503;%%
  %8 citations counted in INSPIRE as of 10 Dec 2014

  
%\cite{Feng:2014bpa}
\bibitem{Feng:2014bpa}
  B.~Feng, D.~Hou and H.~c.~Ren,
  %``(Inverse) Magnetic Catalysis in Bose-Einstein Condensation of Neutral Bound Pairs,''
  arXiv:1412.1647 [cond-mat.quant-gas].
  %%CITATION = ARXIV:1412.1647;%%
  
%\cite{Ferrer:2013noa}
\bibitem{Ferrer:2013noa}
  E.~J.~Ferrer, V.~de la Incera, I.~Portillo and M.~Quiroz,
  %``New look at the QCD ground state in a magnetic field,''
  Phys.\ Rev.\ D {\bf 89}, 085034 (2014)
  [arXiv:1311.3400 [nucl-th]].
  %%CITATION = ARXIV:1311.3400;%%
  %7 citations counted in INSPIRE as of 10 Dec 2014
  
      
\bibitem{Fukushima:2010fe}
  K.~Fukushima, M.~Ruggieri and R.~Gatto,
  %``Chiral magnetic effect in the PNJL model,''
  Phys.\ Rev.\ D {\bf 81}, 114031 (2010)
  [arXiv:1003.0047 [hep-ph]].
  %%CITATION = ARXIV:1003.0047;%%
  %67 citations counted in INSPIRE as of 06 May 2013

  %\cite{Chernodub:2011fr}
\bibitem{Chernodub:2011fr}
  M.~N.~Chernodub and A.~S.~Nedelin,
  %``Phase diagram of chirally imbalanced QCD matter,''
  Phys.\ Rev.\ D {\bf 83}, 105008 (2011)
  [arXiv:1102.0188 [hep-ph]].
  %%CITATION = ARXIV:1102.0188;%%
  %22 citations counted in INSPIRE as of 25 Nov 2014

%\cite{Gatto:2011wc}
\bibitem{Gatto:2011wc}
  R.~Gatto and M.~Ruggieri,
  %``Hot Quark Matter with an Axial Chemical Potential,''
  Phys.\ Rev.\ D {\bf 85}, 054013 (2012)
  [arXiv:1110.4904 [hep-ph]].
  %%CITATION = ARXIV:1110.4904;%%
  %13 citations counted in INSPIRE as of 25 Nov 2014




%\cite{Abelev:2009ac}
\bibitem{Abelev:2009ac}
  B.~I.~Abelev {\it et al.}  [STAR Collaboration],
  %``Azimuthal Charged-Particle Correlations and Possible Local Strong Parity Violation,''
  Phys.\ Rev.\ Lett.\  {\bf 103}, 251601 (2009)
  [arXiv:0909.1739 [nucl-ex]].
  %%CITATION = ARXIV:0909.1739;%%
  %204 citations counted in INSPIRE as of 09 Apr 2014

\bibitem{Abelev:2009ad}
  B.~I.~Abelev {\it et al.}  [STAR Collaboration],
  %``Observation of charge-dependent azimuthal correlations and possible local strong parity violation in heavy ion collisions,''
  Phys.\ Rev.\ C {\bf 81}, 054908 (2010).
 % [arXiv:0909.1717 [nucl-ex]].
  %%CITATION = ARXIV:0909.1717;%%
  %127 citations counted in INSPIRE as of 27 Apr 2013

\bibitem{Abelev:2012pa}
  B.~Abelev {\it et al.}  [ALICE Collaboration],
  %``Charge separation relative to the reaction plane in Pb-Pb collisions at $\sqrt{s_{NN}}= 2.76$ TeV,''
  Phys.\ Rev.\ Lett.\  {\bf 110}, 012301 (2013).
%[arXiv:1207.0900 [nucl-ex]].  %%CITATION = ARXIV:1207.0900;%%  %29 citations counted in INSPIRE as of 27 Apr 2013
%\cite{Abelev:2009ac}


%\cite{Buividovich:2009wi}  {Buividovich:2011cv}
\bibitem{Buividovich:2009wi}
  P.~V.~Buividovich, M.~N.~Chernodub, E.~V.~Luschevskaya and M.~I.~Polikarpov,
  %``Numerical evidence of chiral magnetic effect in lattice gauge theory,''
  Phys.\ Rev.\ D {\bf 80}, 054503 (2009)
  [arXiv:0907.0494 [hep-lat]].
  %%CITATION = ARXIV:0907.0494;%%
  %146 citations counted in INSPIRE as of 20 Jun 2014



%\cite{Ilgenfritz:1988dh}
\bibitem{Ilgenfritz:1988dh}
  E.~-M.~Ilgenfritz and E.~V.~Shuryak,
  %``Chiral Symmetry Restoration at Finite Temperature in the Instanton Liquid,''
  Nucl.\ Phys.\ B {\bf 319}, 511 (1989).
  %%CITATION = NUPHA,B319,511;%%
  %37 citations counted in INSPIRE as of 09 Apr 2014


%\cite{Ilgenfritz:1994nt}
\bibitem{Ilgenfritz:1994nt}
  E.~-M.~Ilgenfritz and E.~V.~Shuryak,
  %``Quark induced correlations between instantons drive the chiral phase transition,''
  Phys.\ Lett.\ B {\bf 325}, 263 (1994)
  [hep-ph/9401285].
  %%CITATION = HEP-PH/9401285;%%
  %60 citations counted in INSPIRE as of 09 Apr 2014


%\cite{Schafer:1994nv}
\bibitem{Schafer:1994nv}
  T.~Schafer, E.~V.~Shuryak and J.~J.~M.~Verbaarschot,
  %``The Chiral phase transition and instanton - anti-instanton molecules,''
  Phys.\ Rev.\ D {\bf 51}, 1267 (1995)
  [hep-ph/9406210].
  %%CITATION = HEP-PH/9406210;%%
  %53 citations counted in INSPIRE as of 09 Apr 2014


\bibitem{Schafer:1996wv}
  T.~Schafer and E.~V.~Shuryak,
  %``Instantons in QCD,''
  Rev.\ Mod.\ Phys.\  {\bf 70}, 323 (1998)
  [hep-ph/9610451].
  %%CITATION = HEP-PH/9610451;%%
  %1065 citations counted in INSPIRE as of 21 Oct 2014



%\cite{Zhang:2012rv}
\bibitem{Zhang:2012rv}
  Z.~Zhang,
  %``Correction to the Chiral Magnetic Effect from axial-vector interaction,''
  Phys.\ Rev.\ D {\bf 85}, 114028 (2012)
  [arXiv:1201.0422 [hep-ph]].
  %%CITATION = ARXIV:1201.0422;%%
  %1 citations counted in INSPIRE as of 09 Apr 2014


%\cite{Kuzmin:1985mm}
%\bibitem{Kuzmin:1985mm}
 % V.~A.~Kuzmin, V.~A.~Rubakov and M.~E.~Shaposhnikov,
  %``On the Anomalous Electroweak Baryon Number Nonconservation in the Early Universe,''
 % Phys.\ Lett.\ B {\bf 155}, 36 (1985).
  %%CITATION = PHLTA,B155,36;%%
  %1918 citations counted in INSPIRE as of 09 Apr 2014


%\cite{Arnold:1987mh}
%\bibitem{Arnold:1987mh}
%  P.~B.~Arnold and L.~D.~McLerran,
  %``Sphalerons, Small Fluctuations and Baryon Number Violation in Electroweak Theory,''
 % Phys.\ Rev.\ D {\bf 36}, 581 (1987).
  %%CITATION = PHRVA,D36,581;%%
  %550 citations counted in INSPIRE as of 09 Apr 2014


%\cite{Arnold:1987zg}
%\bibitem{Arnold:1987zg}
%  P.~B.~Arnold and L.~D.~McLerran,
  %``The Sphaleron Strikes Back,''
%  Phys.\ Rev.\ D {\bf 37}, 1020 (1988).
  %%CITATION = PHRVA,D37,1020;%%
  %334 citations counted in INSPIRE as of 09 Apr 2014


%\cite{Shaposhnikov:1987tw}
%\bibitem{Shaposhnikov:1987tw}
 % M.~E.~Shaposhnikov,
  %``Baryon Asymmetry of the Universe in Standard Electroweak Theory,''
%  Nucl.\ Phys.\ B {\bf 287}, 757 (1987).
  %%CITATION = NUPHA,B287,757;%%
  %484 citations counted in INSPIRE as of 09 Apr 2014


%\cite{McLerran:1990de}
%\bibitem{McLerran:1990de}
 % L.~D.~McLerran, E.~Mottola and M.~E.~Shaposhnikov,
  %``Sphalerons and Axion Dynamics in High Temperature {QCD},''
 % Phys.\ Rev.\ D {\bf 43}, 2027 (1991).
  %%CITATION = PHRVA,D43,2027;%%
  %71 citations counted in INSPIRE as of 09 Apr 2014


%\cite{Giudice:1993bb}
%\bibitem{Giudice:1993bb}
%  G.~F.~Giudice and M.~E.~Shaposhnikov,
  %``Strong sphalerons and electroweak baryogenesis,''
 % Phys.\ Lett.\ B {\bf 326}, 118 (1994)
 % [hep-ph/9311367].
  %%CITATION = HEP-PH/9311367;%%
  %63 citations counted in INSPIRE as of 09 Apr 2014


%\cite{Shuryak:2002qz}
%\bibitem{Shuryak:2002qz}
%  E.~Shuryak and I.~Zahed,
  %``Prompt quark production by exploding sphalerons,''
 % Phys.\ Rev.\ D {\bf 67}, 014006 (2003)
 % [hep-ph/0206022].
  %%CITATION = HEP-PH/0206022;%%
  %22 citations counted in INSPIRE as of 09 Apr 2014

%\cite{Nambu:1961tp}
\bibitem{Nambu:1961tp}
  Y.~Nambu and G.~Jona-Lasinio,
  %``Dynamical Model of Elementary Particles Based on an Analogy with Superconductivity. 1.,''
  Phys.\ Rev.\  {\bf 122}, 345 (1961).
  %%CITATION = PHRVA,122,345;%%
  %3906 citations counted in INSPIRE as of 09 Apr 2014

%\cite{Nambu:1961fr}
\bibitem{Nambu:1961fr}
  Y.~Nambu and G.~Jona-Lasinio,
  %``Dynamical Model Of Elementary Particles Based On An Analogy With Superconductivity. Ii,''
  Phys.\ Rev.\  {\bf 124}, 246 (1961).
  %%CITATION = PHRVA,124,246;%%
  %1866 citations counted in INSPIRE as of 09 Apr 2014

%\cite{Bernard:1987gw}
    \bibitem{Bernard:1987gw}
    V.~Bernard, R.~L.~Jaffe and U.~G.~Meissner,
    %``Flavor Mixing Via Dynamical Chiral Symmetry Breaking,''
    Phys.\ Lett.\ B {\bf 198}, 92 (1987).


%\cite{Bernard:1987sg}
    \bibitem{Bernard:1987sg}
    V.~Bernard, R.~L.~Jaffe and U.~G.~Meissner,
    %``Strangeness Mixing and Quenching in the Nambu-Jona-Lasinio Model,''
    Nucl.\ Phys.\ B {\bf 308}, 753 (1988).

%\cite{Klimt:1989pm}
\bibitem{Klimt:1989pm}
  S.~Klimt, M.~F.~M.~Lutz, U.~Vogl and W.~Weise,
  %``GENERALIZED SU(3) NAMBU-JONA-LASINIO MODEL. Part. 1. MESONIC MODES,''
  Nucl.\ Phys.\ A {\bf 516}, 429 (1990).
  %%CITATION = NUPHA,A516,429;%%
  %230 citations counted in INSPIRE as of 09 Apr 2014

%\cite{Vogl:1991qt}
\bibitem{Vogl:1991qt}
  U.~Vogl and W.~Weise,
  %``The Nambu and Jona Lasinio model: Its implications for hadrons and nuclei,''
  Prog.\ Part.\ Nucl.\ Phys.\  {\bf 27}, 195 (1991).
  %%CITATION = PPNPD,27,195;%%
  %583 citations counted in INSPIRE as of 09 Apr 2014

%\cite{Lutz:1992dv}
\bibitem{Lutz:1992dv}
  M.~F.~M.~Lutz, S.~Klimt and W.~Weise,
  %``Meson properties at finite temperature and baryon density,''
  Nucl.\ Phys.\ A {\bf 542}, 521 (1992).
  %%CITATION = NUPHA,A542,521;%%
  %148 citations counted in INSPIRE as of 14 Nov 2014


%\cite{Klevansky:1992qe}
\bibitem{Klevansky:1992qe}
  S.~P.~Klevansky,
  %``The Nambu-Jona-Lasinio model of quantum chromodynamics,''
  Rev.\ Mod.\ Phys.\  {\bf 64}, 649 (1992).
  %%CITATION = RMPHA,64,649;%%
  %1005 citations counted in INSPIRE as of 09 Apr 2014


%\cite{Hatsuda:1994pi}
\bibitem{Hatsuda:1994pi}
  T.~Hatsuda and T.~Kunihiro,
  %``QCD phenomenology based on a chiral effective Lagrangian,''
  Phys.\ Rept.\  {\bf 247}, 221 (1994)
  [hep-ph/9401310].
  %%CITATION = HEP-PH/9401310;%%
  %1080 citations counted in INSPIRE as of 09 Apr 2014


%\cite{Buballa:2003qv}
\bibitem{Buballa:2003qv}
  M.~Buballa,
  %``NJL model analysis of quark matter at large density,''
  Phys.\ Rept.\  {\bf 407}, 205 (2005)
  [hep-ph/0402234].
  %%CITATION = HEP-PH/0402234;%%
  %643 citations counted in INSPIRE as of 14 Nov 2014





%\cite{'tHooft:1976fv}
\bibitem{'tHooft:1976fv}
  G.~'t Hooft,
  %``Computation of the Quantum Effects Due to a Four-Dimensional Pseudoparticle,''
  Phys.\ Rev.\ D {\bf 14}, 3432 (1976)
  [Erratum-ibid.\ D {\bf 18}, 2199 (1978)].
  %%CITATION = PHRVA,D14,3432;%%
  %3263 citations counted in INSPIRE as of 09 Apr 2014

\bibitem{Shifman:1979uw}
  M.~A.~Shifman, A.~I.~Vainshtein and V.~I.~Zakharov,
  %``Instanton Density in a Theory with Massless Quarks,''
  Nucl.\ Phys.\ B {\bf 163}, 46 (1980).
  %%CITATION = NUPHA,B163,46;%%
  %200 citations counted in INSPIRE as of 02 Sep 2014

%\bibitem{Nowak:1988bh}
%  M.~A.~Nowak, J.~J.~M.~Verbaarschot and I.~Zahed,
%  %``Flavor Mixing in the Instanton Vacuum,''
%  Nucl.\ Phys.\ B {\bf 324}, 1 (1989).
%  %%CITATION = NUPHA,B324,1;%%











\end{thebibliography}
\end{document}